\documentclass{elsarticle}
\usepackage{graphicx}
\usepackage{amssymb}
\usepackage{amsthm}

\newcommand{\beq}{\begin{eqnarray}}
\newcommand{\eeq}{\end{eqnarray}}

\begin{document}

\begin{frontmatter}

\title{The string of variable density: further results}
\author{Paolo Amore}
\ead{paolo.amore@gmail.com}
\address{Facultad de Ciencias, CUICBAS, Universidad de Colima,\\
Bernal D\'{i}az del Castillo 340, Colima, Colima, Mexico}

\begin{abstract}
We analyze the problem of calculating the solutions and the spectrum of a string with arbitrary density
and fixed ends. We build a perturbative scheme which uses a basis of WKB-type functions and
obtain explicit expressions for the eigenvalues and eigenfunctions of the string. Using this approach we show
that it is possible to derive the asymptotic (high energy) behavior of the string, obtaining explicit
expressions for the first three coefficients (the first two can also be obtained with the WKB method).
Finally using an iterative approach we also obtain analytical expressions for the low energy behavior of 
the eigenvalues and eigenfunctions of a string with rapidly oscillating density, recovering 
(in a simpler way) results in the literature.
\end{abstract}


\end{frontmatter}


\section{Introduction}
\label{sec:intro}

In this paper we consider the problem of calculating the normal modes of a string with arbitrary density and finite
length. In a recent paper, ref.~\cite{Amore10b}, we have considered the same problem, and we have
proved the following results: first, that it is possible to apply iterative techniques and obtain increasingly refined
approximations to selected normal modes of the string, starting from a particular ansatz; second, that it is possible to
use a perturbative approach, corresponding to an expansion in the small inhomogeneities of the string, and thus prove that
to second order in this expansion the perturbative results reproduce the asymptotic high energy behavior obtained with the 
WKB method. This result, which is not a--priori obvious, depends crucially on the dimensionality of the problem,
since in one dimension the perturbative contributions depending on off-diagonal terms are not neglegible in the limit
of asymptotically large energies (as it seems to be the case in more than one dimension).

The purpose of this work is twofold: on one hand, to devise an alternative perturbative approach, which is applicable to
strings of arbitrary density (including highly inhomoegeneous cases); on the other hand to show that the iterative theorems of 
ref.~\cite{Amore10b}, may be used to obtain extremely precise (numerical and analytical) results when applied to 
non--trivial problems.

The paper is organized as follows: in Section \ref{sec:dpt} we review the {\sl density perturbation theory} (DPT) of ref.~\cite{Amore10b}; 
in Section \ref{sec:wkbpt} we devise an alternative perturbative method which
uses a basis built out of WKB solutions and derive explicit analytical expressions for the first three coefficients
of the asymptotic expansion for the energy: while the first two coefficients were already known (they can be obtained 
with the standard WKB method), the third coefficient is new (to the best of our knowledge) and it takes the form of an
infinite series of divergent series (each of which can be Borel resummed); in Section \ref{sec:iwkbpt} we describe 
a modified version of this alternative perturbative method, which uses WKB-type solutions, corresponding to non--physical
densities and which may provide a better description of selected modes of the string; in Section \ref{sec:iterative} we
briefly review the iterative theorems of ref.~\cite{Amore10b}; in Section \ref{sec:applications}
we discuss several applications of the methods of previous sections: in particular we test with high precision the 
analytical formula for the third asymptotic coefficient, and we apply the iterative methods to obtain the asymptotic
behavior of the low energy spectrum of a string with highly oscillating density, thus reproducing results 
obtained in the literature~\cite{CZ00b}; finally in Section \ref{sec:conclusions} we draw our conclusions.

\section{Density perturbation theory}
\label{sec:dpt}

The starting point of our discussion is the Helmholtz equation 
\beq
- \frac{d^2\Psi_n(x)}{dx^2} = E_n \rho(x) \Psi_n(x)
\label{sec:dpt:1}
\eeq
which describes the vibrations of a string of density $\rho(x)$. We assume that the string is centered in the origin and that it has 
a length $2L$. The solutions $\Psi_n(x)$  are orthogonal~\cite{BO78}
\beq
\int_{-L}^{+L} \Psi_n(x) \Psi_m(x) \rho(x) dx = \delta_{mn}
\label{sec:dpt:2}
\eeq
and obey specific boundary conditions at the border. In this paper we consider only
Dirichlet boundary conditions, i.e. $\Psi_n(\pm L)= 0$.

Equation (\ref{sec:dpt:1}) can be cast into a more convenient form by introducing the functions 
$\Phi_n(x) \equiv \sqrt{\rho(x)} \Psi_n(x)$ and writing it as
\beq
\frac{1}{\sqrt{\rho(x)}} \left[ - \frac{d^2}{dx^2} \right] \frac{1}{\sqrt{\rho(x)}} \ \Phi_n(x) = E_n \Phi_n(x) \ .
\label{sec:dpt:3}
\eeq

The operator $\hat{O} \equiv \frac{1}{\sqrt{\rho(x)}} \left[ - \frac{d^2}{dx^2} \right] \frac{1}{\sqrt{\rho(x)}}$ has a manifestly
hermitian form and its eigenvalues and eigenfunctions are directly related to those of eq.~(\ref{sec:dpt:1}). 
The results of refs.\cite{Amore10b,Amore10,Amore10c} are obtained using this symmetrized form for the differential operator 
of the inhomogeneous Helmholtz equation.

We now assume that the density $\rho(x)$ of the string may be expressed as
\beq
\rho(x) =  \rho_0 \left( 1 + \eta \ \delta\rho(x) \right) \ ,
\label{sec:dpt:4}
\eeq 
where $\left|\delta\rho(x)\right| \ll 1$ and $\rho(x) >0$ for $x \in (-L,L)$. $\eta$ is a non-physical parameter used 
to allow power-counting (eventually this parameter is set to one). 
Under these assumptions $\delta\rho$ represents a perturbation and therefore one may use perturbation theory.

We expand the operator $\hat{O}$ in a power series
\beq
\hat{O} = \sum_{n=0}^\infty \hat{O}_n \eta^n \ .
\label{sec:dpt:5}
\eeq
where~\cite{Amore10b,Amore10}
\beq
\label{sec:dpt:6a}
\rho_0 \hat{O}_0 &=& -\frac{d^2}{dx^2} \ , \\
\label{sec:dpt:6b}
\rho_0  \hat{O}_1 &=& - \frac{1}{2} \left[ \delta\rho \left(-\frac{d^2}{dx^2}\right) + \left(-\frac{d^2}{dx^2}\right) 
\delta\rho\right] \ , \\
\label{sec:dpt:6c}
\rho_0  \hat{O}_2 &=& \frac{1}{8} \left[ 2 \delta\rho \left(-\frac{d^2}{dx^2}\right) \delta\rho + 3 \delta\rho^2 
\left(-\frac{d^2}{dx^2}\right) + 3 \left(-\frac{d^2}{dx^2}\right) \delta\rho^2 \right]  \ ,  \\
\rho_0 \hat{O}_3   &=&  - \frac{3}{16} \left[  \delta\rho^2 \left(-\frac{d^2}{dx^2}\right) \delta\rho +  
\delta\rho \left(-\frac{d^2}{dx^2}\right) \delta\rho^2 \right]  \nonumber \\
\label{sec:dpt:6d}
&-& \frac{5}{16} \left[  \delta\rho^3 \left(-\frac{d^2}{dx^2}\right)  +   \left(-\frac{d^2}{dx^2}\right) \delta\rho^3 \right]   \ . \\
\dots && \nonumber
\eeq

As standard in perturbation theory we assume that the eigenvalues and eigenfunctions of eq.~(\ref{sec:dpt:3}) can be expressed as power
series of the perturbation density $\delta\rho$, i.e. that
\beq
E_n = \sum_{k=0}^\infty E_n^{(k)} \eta^k \  
\label{sec:dpt:7}
\eeq
and
\beq
\Phi_n(x) =  \sum_{k=0}^\infty \Phi_n^{(k)}(x) \eta^k \ .
\label{sec:dpt:8}
\eeq

Clearly $E_n^{(0)}$  and $\Phi_n^{(0)}(x)$ are the energies and the solutions for a uniform string of density $\rho_0$; we 
also call  $\epsilon_n$ the energies of a uniform string of unit density and use the Dirac bra-ket notation to represent
the modes.

Readers familiar with perturbation theory for stationary states in Quantum Mechanics, i.e. with the Rayleigh-Schr\"odinger 
perturbation theory (RSPT), may easily spot the similarities and the differences between the typical implementation of RSPT 
and the present case: the approach that we follow  here is the standard one used in the treatment of time independent 
perturbations in quantum mechanical problems, although in a typical quantum mechanical problem the Hamiltonian is polynomial 
in the coupling (most likely a linear function of $g$). In the present case the perturbation is the density 
$\delta\rho(x)$ and it appears in a non-polynomial form in the symmetrized operator of eq.(\ref{sec:dpt:4}).

As a result, at a given order in $\eta$ there will be contributions stemming from the operators of different orders $\hat{O}_n$ 
allowed to contribute to that given order and therefore there will be a proliferation of terms as higher orders in $\eta$ are considered.

The first few corrections to the energies of the string have been worked out explicitly in ref.~\cite{Amore10b,Amore10}
and read:
\beq
\label{pt11b}\rho_0 \ E_n^{(0)} &=& \epsilon_n \\
\label{pt12}\rho_0 \ E_n^{(1)} &=& - \epsilon_n \langle n | \delta\rho | n \rangle  \\
\label{pt13}\rho_0 \ E_n^{(2)} &=& \epsilon_n \langle n | \delta\rho | n \rangle^2 
+ \epsilon_n^2 \sum_{k \neq n} \frac{\langle n | \delta\rho | k \rangle^2 }{\epsilon_n-\epsilon_k} \\
\rho_0 \ E_n^{(3)} &=& - \epsilon_n \langle n | \delta\rho | n \rangle^3 + \epsilon_n^3 \langle n | \delta\rho | n \rangle
\sum_{k \neq n}  \frac{\langle n | \delta\rho | k \rangle^2}{\omega_{nk}^2} \nonumber \\
&-& 3 \epsilon_n^2  \langle n | \delta\rho | n \rangle \sum_{k \neq n}  
\frac{\langle n | \delta\rho | k \rangle^2}{\omega_{nk}} \nonumber \\
\label{pt14}&-& \epsilon_n^3 \sum_{k\neq n} \sum_{m \neq n} 
\frac{\langle n | \delta\rho | k \rangle \langle k | \delta\rho | m \rangle 
\langle m | \delta\rho | n \rangle }{\omega_{nk}\omega_{nm}}  \ ,
\eeq
where $\epsilon_n = n^2\pi^2/\rho_0$ and $\omega_{nk} \equiv \epsilon_n -\epsilon_k$.

Here the reader will easily recognize that the direct terms at each order, i.e. the terms involving matrix elements
of $\delta\rho$ between the same states, appear to originate from a geometric series. These terms may be resummed into
\beq
E_n |^{(direct)} \approx \frac{\epsilon_n}{\langle n | \rho(x) | n \rangle}
\eeq
which for $n\rightarrow \infty$ would lead to an asymptotic behaviour
\beq
\lim_{n\rightarrow \infty}E_n |^{(direct)} \approx \frac{\epsilon_n}{\int_{-L}^{+L} \rho(x) dx} \ .
\eeq

In two dimensions, when $\rho(x)$ is the density obtained from a conformal map which transforms an arbitrary 
domain into a square, the denominator of this equation provides the area of original domain, thus leading to a
alternative derivation of Weyl's law, based on perturbation theory~\cite{Amore10c}. The validity of this result however 
requires that the direct terms dominate asymptotically the terms in the perturbation series which mix the different
states. A specific case discussed in ref.~\cite{Amore10} concerning the small deformations of a square, has been proved to
fulfill this requirement to second order.

In one dimension the situation is  different:  for this case we have proved in ref.~\cite{Amore10b} that 
for $n \rightarrow \infty$
\beq
\rho_0 \ \frac{E_n^{(2)}}{\epsilon_n} &\approx& \frac{1}{4} \langle  \sigma^2\rangle  + \frac{3}{4} \langle \sigma \rangle^2 \ .
\label{pert2}
\eeq

This equation is consistent with the asymptotic behaviour which is obtained using the WKB method~\cite{BO78},
\beq
E_n \approx \frac{\epsilon_n}{\langle \sqrt{\rho(x)} \rangle^2} \ ,
\label{weyl1d}
\eeq
where $\langle \sqrt{\rho} \rangle \equiv \frac{1}{2L} \int_{-L}^L \sqrt{\rho(x)} dx $.

One could regard this equation as {\sl Weyl's law for an inhomogeneous string}.

We will now apply the variational principle to obtain useful upper bounds to the energy of the fundamental mode of
a string with arbitrary density. First we observe that the spectrum of the hermitian operator $\hat{O}$ corresponding
to this string is positive and bounded from below; if we let $\Psi(x)$ be an arbitrary function obeying Dirichlet 
boundary conditions at the ends of the string then the expectation value of $\hat{O}$ in this state is an upper bound 
to the exact energy of the fundamental mode:
\beq
\langle \hat{O} \rangle_{\Psi} \geq E_1  \ .
\eeq
The proof of this statement is straightforward once $\Psi(x)$ is expanded in the basis formed by the eigenstates
of $\hat{O}$. 

Working to lowest order  we use the fundamental mode of a uniform string of density $\rho_0$ as a variational ansatz
\beq
\Psi(x) = \sqrt{\frac{1}{L}} \sin \left[ \pi \frac{(x+L)}{2L} \right] 
\eeq
and obtain the bound
\beq
\langle \hat{O} \rangle_{\Psi} &=& - \int_{-L}^{+L} \frac{\pi ^2 \cos ^2\left(\frac{\pi  x}{2 L}\right)}{4 L^3 \rho (x)} dx + 
\int_{-L}^{+L}  \frac{\pi  \sin \left(\frac{\pi  x}{2 L}\right) \cos \left(\frac{\pi  x}{2 L}\right) \rho'(x)}{2 L^2 \rho (x)^2} dx 
\nonumber \\
&-& \int_{-L}^{+L} \ \frac{\cos ^2\left(\frac{\pi  x}{2 L}\right) \rho ''(x)}{2 L \rho (x)^2} dx 
+ \int_{-L}^{+L} \ \frac{3 \cos ^2\left(\frac{\pi  x}{2 L}\right) \rho '(x)^2}{4 L\rho (x)^3} dx \geq E_1 \ .
\eeq
This bound is expected to be useful only for slightly inhomogeneous strings.

\section{WKB perturbation theory}
\label{sec:wkbpt}

We now describe a different approach to the problem of a string of variable  density. This approach has been first 
used by Lindblom and Robiscoe in ref.~\cite{Lindblom91}, where the inhomogeneous Helmholtz equation was cast in the form 
of an equivalent Schr\"odinger equation on the finite domain and approximate expressions for the eigenvalues of 
eq.~(\ref{sec:dpt:1}) were obtained. 

The path that we follow here is similar to that of  ref.~\cite{Lindblom91}, although our primary interest 
lies in the development of a perturbative expansion for the eigenvalues and eigenfunctions of eq.~(\ref{sec:dpt:1}).

We first define the functions
\beq
\phi_n(x) \equiv \sqrt{\frac{2}{\sigma(L)}} \ \rho(x)^{1/4} \ \sin \left[n \pi  \frac{\sigma(x)}{\sigma(L)}\right] 
\ \ \ , \ \ \  n=1,2,\dots 
\label{sec:wkbpt:1}
\eeq
where 
\beq
\sigma(x) \equiv \int_{-L}^{x} \sqrt{\rho(y)}  dy   \ .
\label{sec:wkbpt:2}
\eeq

After making the change of variable $u = \sigma(x)/\sigma(L)$ it is straightforward to see that that these functions are 
orthonormal on $[-L,L]$,  i.e. that 
\beq
\int_{-L}^{+L} \phi_n(x) \phi_m(x)  dx  = \delta_{nm}  \ .
\label{sec:wkbpt:3}
\eeq
This should not be a surprise since the change of variable relates these functions to the eigenfunctions of a uniform 
string. Moreover, the $\phi_n(x)$ are closely related to the WKB functions (see for example eq.(10.1.31) of 
Bender and Orszag, ref.~\cite{BO78}, at pag.490).

Let us now consider the action of the symmetrized differential operator $\hat{O}$ which has been introduced in the 
previous section on any of the functions given above; with elementary algebra we see that
\beq
\hat{O} \phi_n(x) = \left[ \frac{n^2\pi^2}{\sigma(L)^2}  +\frac{4 \rho(x) \rho''(x)-5 \rho'(x)^2}{16 \rho(x)^3} 
\right] \phi_n(x) \ .
\label{sec:wkbpt:4}
\eeq

For the highly excited states ($n \gg 1$) the second term in the parenthesis is neglegible with respect to the first
and the $\phi_n(x)$ become increasingly good approximations to the exact eigenfunctions of $\hat{O}$,  as usual
in the WKB approximation.

The $\phi_n$ are eigenfunctions of the hermitian operator $\hat{Q}$: 
\beq
\hat{Q} \phi_n(x) \equiv \left[\hat{O} - \frac{4 \rho(x) \rho''(x)-5 \rho'(x)^2}{16 \rho(x)^3}\right] \phi_n(x) 
= \frac{n^2\pi^2}{\sigma(L)^2} \phi_n(x) \ ,
\label{sec:wkbpt:6}
\eeq
whose eigenvalues coincide with the leading WKB expression for the eigenvalues of eq.~(\ref{sec:dpt:1}).

In special cases the $\phi_n(x)$ provide the {\sl exact solutions} to the full problem: this happens when the density 
is such that 
\beq
\frac{4 \rho(x) \rho''(x)-5 \rho'(x)^2}{16 \rho(x)^3} = \kappa \ ,
\label{sec:wkbpt:7}
\eeq
where $\kappa$ is a constant. 

The general solution to this ordinary differential equation has the form
\beq
\rho(x) = \frac{256 c_1^2}{\left(c_1^2 \left(c_2+x\right){}^2-256 \kappa \right){}^2} \ ,
\label{sec:wkbpt:8}
\eeq
where $c_1$ and $c_2$ are integration constants; the density discussed by Borg in \cite{Borg46}
is a special case corresponding to 
\beq
c_1 = \frac{16\alpha^2}{1+\alpha} \ \ , \ \ c_2 &=& L + \frac{1}{\alpha} \ \ , \ \ \kappa = 0  \ .
\label{sec:wkbpt:9}
\eeq

With the purpose of implementing a perturbative expansion we define the operator
\beq
\hat{O}_\eta \equiv \hat{Q} + \eta \ \frac{4 \rho(x) \rho''(x)-5 \rho'(x)^2}{16 \rho(x)^3} 
\equiv \hat{Q} + \eta V(\sigma(x)) \ ,
\label{sec:wkbpt:10}
\eeq
where $\eta$ once again is a non-physical parameter used for power counting ($\hat{O}_1 = \hat{O}$) 
and $V(\sigma(x))$ is a potential defined in terms of the density. 
We assume that the eigenvalues and eigenfunctions of $\hat{O}$ can be expressed as power
series in $\eta$, as done before in eqs. (\ref{sec:dpt:7}) and (\ref{sec:dpt:8}) for the DPT. 
Unlike for the DPT we notice that here the density $\rho(x)$ is not bound to be a perturbation 
and that $\hat{O}_{\eta}$ is an operator which depends linearly on $\eta$. 
Because of this last property the corrections for the energies of the string in this approach have a 
simpler form than in those in the DPT. 

We write explicitly the first few terms:
\beq
\label{sec:wkbpt:11a}
E_n^{(0)} &=& \frac{n^2\pi^2}{\sigma(L)^2} \ , \\
\label{sec:wkbpt:11b}
E_n^{(1)} &=& \langle n | V | n \rangle \ , \\ 
\label{sec:wkbpt:11c}
E_n^{(2)} &=& \frac{\sigma(L)^2}{\pi^2} \sum_{k\neq n} \frac{|\langle n | V | k \rangle|^2}{(n^2-k^2)} \\
\label{sec:wkbpt:11d}
E_n^{(3)} &=& \frac{\sigma(L)^4}{\pi^4} \sum_{k\neq n}\sum_{l\neq n} \frac{ \langle n | V | k \rangle \langle k|V| l\rangle 
\langle l | V | n\rangle}{(n^2-k^2)(n^2-l^2)}  \nonumber \\
&-& \frac{\sigma(L)^4}{\pi^4}  \langle n | V | n \rangle  
\sum_{k\neq n} \frac{ \langle n | V | k \rangle^2}{(n^2-k^2)^2}  \\
\dots &=& \dots \nonumber 
\eeq

Using eq.~(\ref{sec:wkbpt:10}) we may cast these expressions directly in terms of the operator $\hat{O}$, taking into 
account that $\langle n | \hat{O} | k \rangle = \langle n | V | k \rangle$ for $k\neq n$.
Notice that the second order correction to the energy of the fundamental mode is always negative as in the standard RSPT.

Similarly for the eigenfunctions we may write 
\beq
\label{sec:wkbpt:12a}
\phi_n^{(0)}(x) &=& \phi_n(x) \\
\label{sec:wkbpt:12b}
\phi_n^{(1)}(x) &=& \frac{\sigma(L)^2}{\pi^2} \sum_{k\neq n} \frac{\langle k | V | n \rangle}{n^2-k^2} \phi_k(x) \\
\phi_n^{(2)}(x) &=& \frac{\sigma(L)^4}{\pi^4} 
\sum_{k\neq n}\sum_{j\neq n}  \frac{\langle k | V | j \rangle \langle j | V | n \rangle}{(n^2-k^2)
(n^2-j^2)} \phi_k(x) \nonumber \\
&-& \frac{\sigma(L)^4}{\pi^4} \langle n | V | n \rangle \sum_{k\neq n} \frac{\langle k | V | n \rangle }{(n^2-k^2)^2} 
\phi_k(x) \\
\phi_n^{(3)}(x) &=& -\frac{\sigma(L)^6}{\pi^6} \langle n | V | n \rangle 
\sum_{k\neq n}\sum_{j\neq n}  \frac{\langle k | V | j \rangle \langle j | V | n \rangle}{(n^2-k^2)^2} \phi_k(x) \nonumber \\
&+&  \frac{\sigma(L)^6}{\pi^6} \langle n | V | n \rangle^2  \sum_{k\neq n} \frac{\langle k | V | n \rangle}{(n^2-k^2)^3} 
\phi_k(x) \nonumber \\
&+& \frac{\sigma(L)^6}{\pi^6} \sum_{k\neq n}\sum_{j\neq n}  \sum_{l\neq n}  
\frac{\langle n | V | l \rangle \langle l | V | n \rangle \langle k | V | j \rangle}{(n^2-k^2)(n^2-j^2)(n^2-l^2)} 
\phi_k(x) \nonumber \\
&-& \frac{\sigma(L)^6}{\pi^6} \langle n | V | n \rangle \sum_{k\neq n}  \sum_{l\neq n}  \frac{\langle n | V | l \rangle 
\langle l | V | n \rangle}{(n^2-k^2)^2 (n^2-l^2)} \phi_k(x) \nonumber \\
&-&  \frac{\sigma(L)^6}{\pi^6} \sum_{k\neq n}  \sum_{j\neq n}  \frac{\langle n | V | l \rangle \langle j | V | n 
\rangle^2}{(n^2-k^2)^2 (n^2-j^2)} \phi_k(x)  \\
\dots &=& \dots \nonumber 
\eeq

The expressions that we have written above are equivalent to those found in ref.~\cite{Lindblom91}; in particular 
eq.(2) of that paper is just the energy of a string calculated to first order in this scheme.

We will now derive asymptotic expressions for the energies of a string of variable density using these 
perturbative formulas and compare them with the asymptotic formulas obtained by Lindblom and Robiscoe.

As we have already mentioned, eq.~(\ref{sec:wkbpt:11a}) is just the leading term for the energy of 
an inhomogeneous string obtained with the standard WKB method~\cite{BO78}. 
The differences with the WKB results appear at first order, in $E_n^{(1)}$; with a simple change of variable we may cast 
eq.~(\ref{sec:wkbpt:11b}) as
\beq
E_n^{(1)} = \frac{2}{\sigma(L)} \ \int_0^{\sigma(L)}  \sin^2 \left[n \pi  \frac{\sigma}{\sigma(L)}\right] V(\sigma(x)) 
d\sigma \ .
\label{sec:wkbpt:13}
\eeq
Eq.(2) of ref.~\cite{Lindblom91} contains this expression.

We may now use the results of  \ref{sec:appendix:1} to write $E_n^{(1)}$ in the equivalent form
\beq
E_n^{(1)} = \langle V \rangle -\sum_{k=0}^\infty \frac{\sigma(L)^{2k+1}}{(2 \pi n)^{2k+2}} (-1)^k 
\left[ V^{(2k+1)}(\sigma(L))  - V^{(2k+1)}(0) \right] \ ,
\label{sec:wkbpt:14}
\eeq
where $\langle V \rangle = \frac{1}{\sigma(L)} \int_0^{\sigma(L)} V(\sigma(x)) d\sigma$ and 
$ V^{(2k+1)}(\sigma) \equiv \frac{d^{2k+1}V(\sigma(x))}{d\sigma^{2k+1}}$. Notice that no approximation is made in 
deriving this expression.

The first term in this expression is independent of $n$ and clearly dominates for $n \gg 1$. We may write it explicitly as
\beq
\langle V \rangle &=& \frac{1}{\sigma(L)} \int_{-L}^{+L} \frac{4 \rho(x) \rho''(x)-5 \rho'(x)^2}{16 \rho(x)^{5/2}} dx 
\label{sec:wkbpt:15}
\eeq
or equivalently as
\beq
\langle V \rangle &=& \frac{\int_{-L}^{+L} \rho(x)^{1/4} \hat{O} \rho(x)^{1/4} dx}{\int_{-L}^{+L} \rho(x)^{1/2} dx } \ .
\eeq

This term agrees with the first correction to the energy that can be obtained within the WKB approximation (in the 
WKB approximation however the "quantization" of energies follows from enforcing the boundary conditions at one end 
of the string) and corresponds to the coefficient $A^{(3)}$ of eq.(25) in ref.\cite{Lindblom91}.

We now consider the contributions of order $1/n^2$, which originate both from $E_n^{(1)}$ and $E_n^{(2)}$. 
We  write eq.(\ref{sec:wkbpt:14}) to this order as
\beq
E_n^{(1)} = \langle V \rangle - \frac{\sigma(L)}{(2 \pi n)^{2}} \left[ V^{(1)}(\sigma(L))  - V^{(1)}(0) \right] 
+ O\left[\frac{1}{n^4}\right] \ .
\label{sec:wkbpt:16}
\eeq
where
\beq
V^{(1)}(\sigma(x)) &=& \frac{dV(\sigma(x))}{d\sigma} = \frac{dV(\sigma(x))}{dx} \frac{1}{\sqrt{\rho(x)}} \nonumber \\
&=& \frac{\rho ^{'''}(x)}{4 \rho (x)^{5/2}}+\frac{15 \rho '(x)^3}{16 \rho (x)^{9/2}}-\frac{9 \rho'(x) \rho''(x)}{8 
\rho (x)^{7/2}} \ .
\label{sec:wkbpt:17}
\eeq

A contribution of order $1/n^2$ stems also from $E_n^{(2)}$. Using the expression for the off-diagonal 
matrix element $\langle n | V|k \rangle$ obtained in  \ref{sec:appendix:1} we may write
\beq
E_n^{(2)} &=& \frac{\sigma^2(L)}{\pi^2} \sum_{k\neq n} \frac{\langle n | V|k \rangle^2}{n^2-k^2} \nonumber \\
&=& \frac{\sigma^2(L)}{\pi^2} \sum_{k\neq n} \sum_{j=0}^\infty\sum_{l=0}^\infty (-1)^{j+l} 
\frac{\sigma^{2 (j+l)+2}(L)}{\pi^{2(j+l)+4}} \frac{1}{n^2-k^2} \nonumber \\
&\cdot& \left(\frac{1}{(k-n)^{2j+2}}-\frac{1}{(k+n)^{2j+2}} \right)
\left(\frac{1}{(k-n)^{2l+2}}-\frac{1}{(k+n)^{2l+2}} \right) \nonumber \\
&\cdot& \left[ (-1)^{k+n} V^{(2j+1)}(\sigma(L))-V^{(2j+1)}(0)\right] \nonumber \\
&\cdot& \left[ (-1)^{k+n} V^{(2l+1)}(\sigma(L))-V^{(2l+1)}(0)\right] \nonumber \ ,
\eeq
where no approximation has been made. 

This expression may be cast in the more compact form
\beq
E_n^{(2)} &=& \frac{\sigma^2(L)}{\pi^2} \sum_{j=0}^\infty\sum_{l=0}^\infty (-1)^{j+l} 
\frac{\sigma^{2 (j+l)+2}(L)}{\pi^{2(j+l)+4}} \nonumber \\
&\cdot& \left\{S^{(I)}_{jl}(n)  \left[ V^{(2j+1)}(\sigma(L))V^{(2l+1)}(\sigma(L))
+ V^{(2j+1)}(0) V^{(2l+1)}(0)\right]  \right. \nonumber \\
&-& \left. S^{(II)}_{jl}(n) \left[ V^{(2j+1)}(\sigma(L))V^{(2l+1)}(0)
+ V^{(2j+1)}(0) V^{(2l+1)}(\sigma(L))\right] \right\} \ ,
\label{EN2}
\eeq
where we have defined the series
\beq
S^{(I)}_{jl}(n) &\equiv& \sum_{k\neq n}  \frac{1}{n^2-k^2} \left(\frac{1}{(k-n)^{2j+2}}-\frac{1}{(k+n)^{2j+2}} \right) \nonumber \\
&\cdot& \left(\frac{1}{(k-n)^{2l+2}}-\frac{1}{(k+n)^{2l+2}} \right) \nonumber \\
S^{(II)}_{jl}(n) &\equiv& \sum_{k\neq n}  \frac{(-1)^{k+n}}{n^2-k^2} \left(\frac{1}{(k-n)^{2j+2}}-\frac{1}{(k+n)^{2j+2}} \right) \nonumber \\
&\cdot& \left(\frac{1}{(k-n)^{2l+2}}-\frac{1}{(k+n)^{2l+2}} \right) \nonumber \ .
\eeq

We are interested in calculating the dominant contributions in these expressions for $n \rightarrow \infty$:
\beq
S^{(I)}_{jl}(n) &\approx& \frac{\zeta( 2 (2+j+l))}{2n^2}  + O\left[\frac{1}{n^4}\right] \\
S^{(II)}_{jl}(n)&\approx& \frac{\zeta( 2 (2+j+l))}{n^2}  \left[-\frac{1}{2} + 2^{-2 (2+j+l)}\right]+ 
O\left[\frac{1}{n^3}\right]  \nonumber \\
&=& - \frac{\zeta( 2 (2+j+l))}{2n^2}  + \frac{1}{n^2}
\sum_{k=1}^\infty \frac{1}{(2k)^{2 (2+j+l)}} + O\left[\frac{1}{n^4}\right]  \ .
\eeq

Therefore we may write:
\beq
E_n^{(2)} &=& \frac{\sigma^2(L)}{n^2\pi^2} \sum_{j=0}^\infty\sum_{l=0}^\infty (-1)^{j+l} 
\frac{\sigma^{2 (j+l)+2}(L)}{\pi^{2(j+l)+4}} \nonumber \\
&\cdot& \left\{
\frac{1}{2} \sum_{k=1}^\infty \frac{1}{(k)^{2 (2+j+l)}} 
 \left[ V^{(2j+1)}(\sigma(L)) V^{(2l+1)}(\sigma(L))
+ V^{(2j+1)}(0) V^{(2l+1)}(0) \right. \right. \nonumber \\
&+& \left. \left. V^{(2j+1)}(\sigma(L))V^{(2l+1)}(0)
+ V^{(2j+1)}(0) V^{(2l+1)}(\sigma(L))
\right]  \right. \nonumber \\
&-& \left. 
\sum_{k=1}^\infty \frac{1}{(2k)^{2 (2+j+l)}} 
\left[ V^{(2j+1)}(\sigma(L))V^{(2l+1)}(0)
+ V^{(2j+1)}(0) V^{(2l+1)}(\sigma(L))\right] \right\} \nonumber \\
&+& O\left[\frac{1}{n^4}\right]  \ .
\eeq

Let us define the series~\footnote{We will soon discuss examples where this series is divergent.}
\beq
\mathcal{F}(\upsilon,\sigma_0) \equiv - i \sum_{j=0}^\infty i^{2j+1} \upsilon^{2j+1} 
V^{(2j+1)}(\sigma_0) \ ,
\eeq
which is related to the series
\beq
\mathcal{G}(\upsilon,\sigma_0) &\equiv& - i \sum_{j=0}^\infty \frac{i^{2j+1}}{(2j+1)!} 
\upsilon^{2j+1}  V^{(2j+1)}(\sigma_0) \nonumber \\
&=&  -\frac{i}{2} \left[ V(\sigma_0 + i \upsilon) - V(\sigma_0 - i \upsilon)
\right]
\eeq
by a Borel transform
\beq 
\mathcal{F}(\upsilon,\sigma_0) &=& \int_0^{\infty}  e^{-t} \mathcal{G}(t \upsilon,\sigma_0) dt \nonumber \ .
\eeq

We can therefore cast $E_n^{(2)}$ directly in terms of $\mathcal{F}$ as
\beq
E_n^{(2)} &=& \frac{\sigma(L)^2}{n^2\pi^4} \sum_{k=1}^\infty \frac{1}{2k^2} 
\left\{ \left[ \mathcal{F}\left( \frac{ \sigma(L)}{\pi k} , \sigma(L) \right) + 
\mathcal{F}\left( \frac{ \sigma(L)}{\pi k} , 0 \right)  \right]^2 \right. \nonumber \\
&-& \left.  
\mathcal{F}\left( \frac{ \sigma(L)}{2\pi k} , \sigma(L) \right) 
\mathcal{F}\left( \frac{ \sigma(L)}{2\pi k} , 0 \right) \right\} + O\left[\frac{1}{n^4}\right] \  . 
\eeq
To the best of our knowledge this result was not known.

Therefore, the exact expression for the energy to order $1/n^2$ is 
\beq
E_n &=& \frac{\pi^2}{\sigma(L)^2} n^2 +  \langle V \rangle + \frac{1}{n^2}  \left\{ - \frac{\sigma(L)}{4 \pi^2} 
\left[ V^{(1)}(\sigma(L))  - V^{(1)}(0) \right]  \right. \nonumber \\
&+& \left.  \frac{\sigma(L)^2}{\pi^4} \sum_{k=1}^\infty \frac{1}{2k^2} 
\left\{ \left[ \mathcal{F}\left( \frac{ \sigma(L)}{\pi k} , \sigma(L) \right) + 
\mathcal{F}\left( \frac{ \sigma(L)}{\pi k} , 0 \right)  \right]^2 \right.\right. \nonumber \\
&-& \left.  \left.
\mathcal{F}\left( \frac{ \sigma(L)}{2\pi k} , \sigma(L) \right) 
\mathcal{F}\left( \frac{ \sigma(L)}{2\pi k} , 0 \right) \right\} \right\} 
+ O\left[\frac{1}{n^4}\right] \ .
\eeq
In the next sections we will explicitly check this formula working on specific examples.

To make contact with the results of Ref.~\cite{Lindblom91} we consider a simple approximation to $E_n^{(2)}$,
which is obtained restricting the sums over $j$ and $l$ in eq.(\ref{EN2}) to $j=l=0$~\footnote{Clearly this formula
is not expected to be accurate, unless the contributions to the term which goes like $n^{-2}$ stemming from 
$E_n^{(2)}$ are particularly small compared to the one stemming from $E_n^{(1)}$.}; 
in this case we have
\beq
 S^{(I)}_{00}(n) &=&  \frac{\pi^4}{180 n^2} + O\left[\frac{1}{n^4}\right] \\
 S^{(II)}_{00}(n) &=& - \frac{7}{1440 n^2}  + O\left[\frac{1}{n^4}\right] ,
\eeq
and the energy is approximately given by
\beq
E_n^{(2)} &\approx& \frac{\sigma^4(L)}{\pi^2n^2}  \left\{ \frac{1}{180} \left[ \left(V^{(1)}(\sigma(L))\right)^2
+ \left(V^{(1)}(0)\right)^2\right] \right. \nonumber \\
&+& \left. \frac{7}{720} V^{(1)}(\sigma(L)) V^{(1)}(0) \right\} \ .
\label{en2a}
\eeq

To allow the comparison with the eq.(15) and (20) of \cite{Lindblom91} we write the approximate analytical expression for the 
energies of the inhomogeneous string using eq.~(\ref{en2a}) (i.e. restricting the double series in the expression for the $1/n^2$ coefficient 
inside $E_n^{(2)}$ to the first term):
\beq
E_n &\approx& \frac{\pi^2}{\sigma(L)^2} n^2 +  \langle V \rangle + \frac{1}{n^2}  \left\{ - \frac{\sigma(L)}{4 \pi^2} 
\left[ V^{(1)}(\sigma(L))  - V^{(1)}(0) \right]  \right. \nonumber \\
&+& \left.
\frac{\sigma(L)^4}{\pi^2} \left[ \frac{ V^{(1)}(\sigma(L))^2+ V^{(1)}(0)^2}{180} + \frac{7}{720}  V^{(1)}(\sigma(L))  V^{(1)}(0)
\right] +\dots \right\} \nonumber \\
&+& O\left[\frac{1}{n^4}\right] \ .
\eeq

There are some important considerations to make: to start with, it should be easy to convince oneself that the asymptotic 
expansion of the energy is an even function of $n$ and that contributions to a given order $1/n^{2p}$ can only come from the 
first $p$ perturbative orders. 
This happens because the non-diagonal matrix elements go to zero as $n \rightarrow \infty$ and because the WKBPT contributions 
to a given order $p$ contain $p$ of these matrix elements (an example to this is given by the calculation
that we have just performed, where the corrections to the energy decaying as $1/n^2$ only come from $E_n^{(1)}$ and $E_n^{(2)}$).
This means that the expression that we have written above is exact to order $1/n^2$ and that the evaluation of higher order terms 
can be done by using a larger (but finite) number of perturbative terms.

The second point concerns the relation of our expression with the  one obtained by Lindblom and Robiscoe, who estimate the
second order contributions using the mean value theorem for derivatives. The two expressions {\sl are not equivalent} since the results 
of ref.~\cite{Lindblom91} are obtained assuming that the density is a slowly varying function as stated in their eqns.(11) and (12);
we have not made this hypothesis on the density in our calculation. Morevover the term $V(x)$ in the equation (\ref{sec:wkbpt:4}) is always 
a perturbation for $n$ sufficiently large provided that $V(x)$ is not singular on the domain, as previously mentioned. 
It is straightforward to see that our result of eq.(\ref{en2a}) reduces to the result of ref.~\cite{Lindblom91} in the case where 
$V^{(1)}$ is a symmetric  function of $x$: in this case we have that $ V^{(1)}(\sigma(L)) =  V^{(1)}(0)$ and 
$2/180+7/720$ reduces to the factor $1/48$ of ref.~\cite{Lindblom91}.

We will now use the WKBPT to obtain bounds on the energy of the fundamental mode of a string of arbitrary density,
using the solutions in eqs.~(\ref{sec:wkbpt:12a}) and (\ref{sec:wkbpt:12b}) to build a suitable variational ansatz.
As we have seen in the previous section, for an arbitrary (normalized) ansatz $\Psi(x)$, the Rayleigh quotient provides 
an upper bound to the true energy:
\beq
\langle \hat{O} \rangle_{\Psi} \geq E_1 \ . \nonumber
\eeq

In this case we will use the wave function for the fundamental mode obtained using WKBPT to a finite order as a variational ansatz;
the simplest approach corresponds to choosing the lowest order expression as the ansatz
\beq
\Psi(x) = \phi_n^{(0)}(x) \ ,
\eeq
which allows one to obtain the bound
\beq
\langle \hat{O}\rangle_{\Psi} &=& \frac{\pi^2}{\sigma(L)^2} + \frac{2}{\sigma(L)} \ 
\int_{-L}^{+L} \sin^2 \left[\frac{\pi\sigma(x)}{\sigma(L)}\right] \ V(x) \sqrt{\rho(x)} dx \nonumber \\
&\equiv& E_1^{(0)} + E_1^{(1)} \geq E_1 \ ,
\label{bound_wkbpt1}
\eeq
which is just eq.(2) of ref.~\cite{Lindblom91} for $n=1$, although the authors of that paper do not mention this property explicitly.

The first term in this expression is the leading WKB expression, which typically is not very accurate for the fundamental 
mode, and it is always positive; the second term, on the other hand, depends on $V(x)$ and it does not have a definite sign. 
In the WKB method this last term reduces to $\frac{1}{\sigma(L)} \ \int_{-L}^{+L} V(x) \sqrt{\rho(x)} dx$, having substituted the 
$\sin^2\left[\frac{\pi\sigma(x)}{\sigma(L)}\right]$ with its average value on the interval. Unfortunately, the WKB expression
neither provides a bound nor it is accurate, as we will later see in a specific example. 

If we like to obtain a stricter upper bound for the energy of the fundamental mode of our inhomogeneous string we
may use the ansatz (not normalized)
\beq
\Psi(x) = \phi_1^{(0)}(x) + \phi_1^{(1N)}(x) \ ,
\eeq
which corresponds to the solution for the fundamental mode calculated with WKBPT to first order, restricting the sum in 
the first order contribution to the first $N-1$ terms:
\beq
\phi_1^{(1N)}(x) &=& \frac{\sigma(L)^2}{\pi^2} \sum_{k=2}^N \frac{\langle k | V | 1 \rangle}{1-k^2} \phi_k(x) \ .
\eeq

In this case
\beq
\langle \hat{O}\rangle_{\Psi} = \frac{\frac{\pi^2}{\sigma(L)^2} + \langle \phi_1^{(1)} \hat{Q} \phi_1^{(1)}\rangle + 
\langle \Psi V \Psi \rangle}{1 + \langle {\phi_1^{(1)}}^2 \rangle  } \geq E_1 \ ,
\eeq
where
\beq
\langle {\phi_1^{(1)}}^2 \rangle &=& \left[\frac{\sigma(L)}{\pi}\right]^4 \sum_{k=2}^N \frac{\langle 1 | V|k \rangle^2}{(1-k^2)^2} \\
\langle \phi_1^{(1)} \hat{Q} \phi_1^{(1)}\rangle &=& \left[\frac{\sigma(L)}{\pi}\right]^2 \sum_{k=2}^N k^2 
\frac{\langle 1 | V|k \rangle^2}{(1-k^2)^2}  \ , \\
\langle \Psi V \Psi \rangle &=& \langle 1 | V | 1 \rangle + 2 \left[\frac{\sigma(L)}{\pi}\right]^2 \sum_{k=2}^N 
\frac{\langle 1 | V|k \rangle^2}{(1-k^2)} \ , \nonumber \\
&+& \left[\frac{\sigma(L)}{\pi}\right]^4 \sum_{k,l\neq 1}  
\frac{\langle 1 | V|k \rangle \langle 1 | V|l \rangle\langle l | V|k \rangle}{(1-k^2) (1-l^2)}  \ .
\eeq

We will also apply this second bound later to specific examples.

\section{Improved WKB perturbation theory}
\label{sec:iwkbpt}

We will now describe an alternative implementation of WKBPT which is suitable for describing specific parts of the spectrum.

Let us consider an unphysical density $\tilde{\rho}(x)$: by unphysical density we mean a positive function defined on the 
domain of the string ($\tilde{\rho}(x) > 0$ for $|x| \leq L$) and different from the physical density of the string, 
$\tilde{\rho}(x) \neq \rho(x)$. 

We define $\tilde{\phi}_n(x)$ the functions corresponding to eqn. (\ref{sec:wkbpt:1}) for $\rho(x) \rightarrow \tilde{\rho}(x)$
\beq
\tilde{\phi}_n(x) \equiv \sqrt{\frac{2}{\tilde{\sigma}(L)}} \ \tilde{\rho}(x)^{1/4} \ \sin \left[n \pi  
\frac{\tilde{\sigma}(x)}{\tilde{\sigma}(L)}\right] \ \ \ , \ \ \ 
n=1,2,\dots 
\label{sec:iwkbpt:1}
\eeq
where 
\beq
\tilde{\sigma}(x) \equiv \int_{-L}^{x} \sqrt{\tilde{\rho}(y)}  dy   \ .
\label{sec:iwkbpt:2}
\eeq

The $\tilde{\phi}_n(x)$ are eigenfunctions of an operator $\hat{\tilde{Q}}$, obtained form
$\hat{Q}$ with the substitution $\rho(x) \rightarrow \tilde{\rho}(x)$ and they are isospectral with the $\phi_n(x)$ 
if $\tilde{\sigma}(L) = \sigma(L)$.

If we apply $\hat{O}$ to $\tilde{\phi}_n(x)$ we see that 
\beq
\hat{O} \tilde{\phi}_n(x) = \frac{n^2\pi^2}{\tilde{\sigma}(L)^2} \frac{\tilde{\rho}(x)}{\rho(x)} \left[ 1 + O\left(\frac{1}{n}\right)
\right]\tilde{\phi}_n(x)  \ ,
\eeq
i.e. that only the functions corresponding to $\tilde{\rho}(x) = \rho(x)$ are eigensolutions of $\hat{O}$ for 
$n  \rightarrow \infty$. On the other hand, one can pick a specific density $\tilde{\rho}(x)$ so 
that a given state of $\tilde{\hat{Q}}$ is also eigenstate of $\hat{O}$. We can think of this density
as an {\sl effective density}, which will reduce to the physical density for states with $n \rightarrow \infty$.

This density may be obtained as follows: suppose that $\Psi_n(x)$ is the $n^{th}$ eigensolution of $\hat{O}$, and 
let $\tilde{\phi}_n(x)$ be the $n^{th}$ eigensolution of $\tilde{\hat{Q}}$. We want to find the density 
$\tilde{\rho}(x)$ for which
\beq
\Psi_n(x) = \tilde{\phi}_n(x) \ ,
\eeq
which is equivalent to the equation
\beq
\frac{\tilde{\sigma}(x)}{\tilde{\sigma}(L)} - \frac{\sin\left[ \frac{2n\pi\tilde{\sigma}(x)}{\tilde{\sigma}(L)}\right]}{2n\pi} =
\int_{-L}^x \Psi_n(y)^2 dy \ .
\eeq
Once $\tilde{\sigma}(x)$ is known, the density is obtained using the relation $\tilde{\sigma}'(x) = \sqrt{\tilde{\rho}(x)}$.

The procedure outlined above explains how the effective density for a given state of $\hat{O}$ can be obtained once that
the eigenfunction is known explicitly; in a typical situation however the eigenfunctions of $\hat{O}$ are not
known and need to be calculated. In this case one must resort to an alternative procedure.

If we are interested in describing the lowest mode of a string with density $\rho(x)$, we may use the variational principle
to obtain the effective density $\tilde{\rho}(x)$ for this mode: in this case the eigenfunction corresponding to the 
lowest mode of $\tilde{\hat{Q}}$ can be used as a variational ansatz and therefore one can obtain the 
inequality
\beq
\frac{2}{\tilde{\sigma}(L)} \int_{-L}^{+L} \frac{\tilde{\rho}(x)^{1/4}}{\sqrt{\rho(x)}}
\sin \left[\frac{\pi \tilde{\sigma}(x)}{\tilde{\sigma}(L)} \right] \left( - \frac{d^2}{dx^2} \right) 
\frac{\tilde{\rho}(x)^{1/4}}{\sqrt{\rho(x)}} \sin \left[\frac{\pi \tilde{\sigma}(x)}{\tilde{\sigma}(L)} \right] dx  \geq E_1 \ .
\label{bound_iwkbpt}
\eeq

The LHS of this inequality is just the expectation value of $\hat{O}$ in the fundamental mode of a string with density 
$\tilde{\rho}(x)$~\footnote{Notice that $\langle \hat{O}\rangle$ is invariant under the redefinition of the density 
$\tilde{\rho}(x) \rightarrow c \tilde{\rho}(x)$,  where $c$ is a constant.}.
If $\tilde{\rho}(x)$ depends on one or more parameters which can be varied without $\tilde{\rho}(x)$ changing sign on all domain,
then one obtains an optimal approximation to the fundamental mode of $\hat{O}$ in correspondence of an absolute minimum of the LHS
of the inequality.  The density obtained using this values of the parameters approximates the effective density for the 
fundamental mode of $\hat{O}$.

The bound discussed above can still be improved: Gottlieb has shown in ref.~\cite{Gottlieb02} that a string of 
length $2L$ centered in the origin and with density 
\beq
\bar{\rho}(x) = \xi'(x)^2 \ \rho(\xi(x))
\eeq
where $\alpha > -1/2L$ is a real parameter and
\beq
\xi(x) = \frac{1+ 2\alpha L}{1+\alpha (x+L)} (x+L)-L
\eeq
is isospectral to a string of the same length also centered in the origin and with density $\rho(x)$~\footnote{We may easily prove
this property by proving that $V$ is invariant under this transformation.}.
Using this property one may express the bound (\ref{bound_iwkbpt}) for the operator $\bar{\hat{O}} = \frac{1}{\sqrt{\bar{\rho}(x)}}
\left( - \frac{d^2}{dx^2} \right) \frac{1}{\sqrt{\bar{\rho}(x)}}$. By choosing $\alpha$ suitably one may obtain a more accurate bound
for the energy of the fundamental mode of the string.

In the case of arbitrary excited modes of $\hat{O}$ the approach to obtain the effective density for these modes is different.
Let us consider an arbitrary trial function $\Psi_n(x) = \langle x |\Psi_n\rangle$, with $\Psi_n(\pm L) = 0$ and
with $n-1$ nodes in the interval $x \in (-L,L)$. 
We wish to obtain accurate approximations to the eigenfunctions of the hermitian operator 
$\hat{O} = \frac{1}{\sqrt{\rho(x)}} \left(- \frac{d^2}{dx^2}\right) \frac{1}{\sqrt{\rho(x)}}$. For this reason
we define the quantity
\beq
\Sigma_n = \langle \Psi_n | \hat{O}^2 | \Psi_n \rangle - \langle \Psi_n | \hat{O} | \Psi_n \rangle^2 \geq 0 \ ,
\label{eq:Sigma}
\eeq
where the equality holds only when $|\Psi_n\rangle$ is a solution of $\hat{O}$. 

Provided that we vary $\Psi_n(x)$ without changing the number of nodes in $(-L,+L)$, the optimal 
approximation to the $n^{th}$ solution of the string falling in this class of functions will be the one which 
minimizes $\Sigma_n$. We can achieve this goal using a trial wave function corresponding to the $n^{th}$ solution of
a string of arbitrary density $\tilde{\rho}(x)$, which depends on one or more arbitrary parameters:
\beq
\Psi_n(x) = \tilde{\phi}_n(x) =  \sqrt{\frac{2}{\tilde{\sigma}(L)} } \ \tilde{\rho}(x)^{1/4} \ \sin \frac{n \pi \tilde{\sigma}(x)}{\tilde{\sigma}(L)}  \ .
\eeq

By varying the parameters in $\tilde{\rho}(x)$, with $\tilde{\rho}(x)>0$ on the interval, we modify the trial solution, 
without changing the number of nodes and therefore the minimization of the corresponding $\Sigma_n$ will yield
the optimal approximation to the mode that we want to calculate. The corresponding $\tilde{\rho}(x)$ will approximate
the effective density for this mode.

From a computational point of view the calculation of $\Sigma_n$ as given in eq.~(\ref{eq:Sigma}) is not efficient since
the difference of small quantities may lead to large numerical errors and even to negative values of $\Sigma_n$. A more efficient
calculation can be done using the completeness  of the orthonormal basis to which $\tilde{\phi}_n(x)$ belongs and writing
\beq
\Sigma_n = \sum_{k\neq n}^{\infty} \langle n | \hat{O} | k \rangle^2 \ ,
\eeq
where $\langle x | k \rangle = \tilde{\phi}_k(x)$. In this case $\Sigma_n$ is the sum of positive quantities and therefore  
the numerical errors do not affect the sign of $\Sigma_n$. 

Provided that the matrix elements $\langle n | \hat{O} | k\rangle$ fall off sufficiently rapidly away from the principal diagonal, 
the series may be truncated to a finite number of terms, without introducing large errors:
\beq
\Sigma_n^{(m)} \approx \sum_{k_{min},k\neq n}^{k_{max}} \langle n | \hat{O} | k \rangle^2 \ ,
\eeq
where $k_{min} \equiv \max ( k-m,1)$ and $k_{max}=k+m$. 

We are now ready to describe a new perturbation scheme which uses the effective density of a given state of $\hat{O}$ 
and the set of functions related to this density, $\tilde{\phi}_n(x)$, as the "unperturbed" basis.
We write the operator
\beq
\hat{O}_\eta \equiv \tilde{\hat{Q}} + \eta \left[ \hat{O} - \tilde{\hat{Q}} \right] \equiv \hat{\tilde{Q}} + \eta \hat{W} \ , 
\eeq
where $\tilde{\hat{Q}}$ is the operator obtained from the hermitian operator corresponding to a fictitious string
of density $\tilde{\rho}(x)$, 
\beq
\hat{\tilde{O}} \equiv \frac{1}{\sqrt{\tilde{\rho}(x)}}  \left(- \frac{d^2}{dx^2} \right) \frac{1}{\sqrt{\tilde{\rho}(x)}}  \ 
\eeq
by means of the decomposition
\beq
\hat{\tilde{O}} = \hat{\tilde{Q}} + \tilde{V}(x) \ ,
\eeq
with
\beq
\tilde{V}(x) \equiv \frac{4 \tilde{\rho}(x) \tilde{\rho}''(x)-5 \tilde{\rho}'(x)^2}{16 \tilde{\rho}(x)^3}  \ .
\eeq

Notice that when $\tilde{\rho}(x)$ is chosen as a constant density $\rho_0$ 
we have that $\tilde{V} = 0$ and the operator $\hat{\tilde{Q}} =\hat{\tilde{O}}$. In this case one may write 
\beq
\hat{O}_\eta \equiv -\frac{1}{\rho_0} \frac{d^2}{dx^2} + \eta \left[- \frac{1}{\sqrt{\rho(x)}} \frac{d^2}{dx^2} \frac{1}{\sqrt{\rho(x)}}
+ \frac{1}{\rho_0} \frac{d^2}{dx^2}  \right] \ .
\eeq

The Density Perturbation Theory (DPT) discussed early is recovered after expressing $\rho(x) = \rho_0 (1+\delta\rho(x))$,
with $\rho_0$ the average value of $\rho(x)$ over the interval, and expanding 
in powers of $\delta\rho(x)$: therefore DPT may be regarded as a special case of the present scheme.

Going back to the general expression, we point out that $\eta$ is a parameter allowing power-counting and that
$\hat{O}_{\eta=1} = \hat{O}$;  we treat $\hat{W}$ as a perturbation and
obtain the perturbative expressions for the energies and eigenfunctions of a string of density $\rho(x)$ 
as~\footnote{Here we call $| n \rangle$ the eigenstates of $\tilde{\hat{Q}}$, thus implying  
that $\langle x | n \rangle = \tilde{\phi}_n(x)$.}
\beq
\label{sec:wkbpt:21a}
E_n^{(0)} &=& \frac{n^2\pi^2}{\tilde{\sigma}(L)^2} \ , \\
\label{sec:wkbpt:21b}
E_n^{(1)} &=& \langle n | W | n \rangle \ , \\ 
\label{sec:wkbpt:21c}
E_n^{(2)} &=& \frac{\tilde{\sigma}(L)^2}{\pi^2} \sum_{k\neq n} \frac{|\langle n | W | k \rangle|^2}{(n^2-k^2)} \\
\label{sec:wkbpt:21d}
E_n^{(3)} &=& \frac{\tilde{\sigma}(L)^4}{\pi^4} \sum_{k\neq n}\sum_{l\neq n} \frac{ \langle n | W | k \rangle \langle k| W | l\rangle 
\langle l | W | n\rangle}{(n^2-k^2)(n^2-l^2)}  \nonumber \\
&-& \frac{\tilde{\sigma}(L)^4}{\pi^4}  \langle n | W | n \rangle  
\sum_{k\neq n} \frac{ \langle n | W | k \rangle^2}{(n^2-k^2)^2}  \\
\dots &=& \dots \nonumber 
\eeq
and
\beq
\label{sec:wkbpt:22a}
\phi_n^{(0)}(x) &=& \tilde{\phi}_n(x) \\
\label{sec:wkbpt:22b}
\phi_n^{(1)}(x) &=& \frac{\tilde{\sigma}(L)^2}{\pi^2} \sum_{k\neq n} \frac{\langle k | W | n \rangle}{n^2-k^2} \tilde{\phi}_k(x) \\
\phi_n^{(2)}(x) &=& \frac{\tilde{\sigma}(L)^4}{\pi^4} 
\sum_{k\neq n}\sum_{j\neq n}  \frac{\langle k |W| j \rangle \langle j |W| n \rangle}{(n^2-k^2)
(n^2-j^2)} \tilde{\phi}_k(x) \nonumber \\
&-& \frac{\tilde{\sigma}(L)^4}{\pi^4} \langle n |W| n \rangle \sum_{k\neq n} \frac{\langle k |W| n \rangle }{(n^2-k^2)^2} \tilde{\phi}_k(x) \\
\phi_n^{(3)}(x) &=& -\frac{\tilde{\sigma}(L)^6}{\pi^6} \langle n | W | n \rangle 
\sum_{k\neq n}\sum_{j\neq n}  \frac{\langle k | W | j \rangle \langle j | W | n \rangle}{(n^2-k^2)^2} \tilde{\phi}_k(x) \nonumber \\
&+&  \frac{\tilde{\sigma}(L)^6}{\pi^6} \langle n | W | n \rangle^2  \sum_{k\neq n} \frac{\langle k | W | n \rangle}{(n^2-k^2)^3} 
\tilde{\phi}_k(x) \nonumber \\
&+& \frac{\tilde{\sigma}(L)^6}{\pi^6} \sum_{k\neq n}\sum_{j\neq n}  \sum_{l\neq n}  
\frac{\langle n | W | l \rangle \langle l | W | n \rangle \langle k | W | j \rangle}{(n^2-k^2)(n^2-j^2)(n^2-l^2)} \tilde{\phi}_k(x) 
\nonumber \\
&-& \frac{\tilde{\sigma}(L)^6}{\pi^6} \langle n | W | n \rangle \sum_{k\neq n}  \sum_{l\neq n}  \frac{\langle n | W | l \rangle 
\langle l | W | n \rangle}{(n^2-k^2)^2 (n^2-l^2)} \tilde{\phi}_k(x) \nonumber \\
&-&  \frac{\tilde{\sigma}(L)^6}{\pi^6} \sum_{k\neq n}  \sum_{j\neq n}  \frac{\langle n | W | l \rangle 
\langle j | W | n \rangle^2}{(n^2-k^2)^2 (n^2-j^2)} \tilde{\phi}_k(x)  \\ 
\dots &=& \dots \nonumber 
\eeq

Having chosen $\tilde{\rho}(x)$ to approximate the "effective" density for the $n^{th}$ excited state of $\hat{O}$ we expect that this 
perturbation scheme will be helpful to obtain precise approximations to this state and to states which fall near since in this basis 
the $n^{th}$ state is almost decoupled from the remaining states.

We will now discuss the asymptotic behaviour of the energies calculated to first order:
\beq
E_n^{(0)} &+& E_n^{(1)} = \frac{n^2 \pi^2}{\tilde{\sigma}(L)^2}  + \langle n | \hat{W} | n \rangle 
\nonumber \\
&=& \frac{2}{\tilde{\sigma}(L)} \int_{-L}^{+L} \frac{\tilde{\rho}(x)^{1/4}}{\sqrt{\rho(x)}}
\sin \left[\frac{n \pi \tilde{\sigma}(x)}{\tilde{\sigma}(L)} \right] \left( - \frac{d^2}{dx^2} \right) 
\frac{\tilde{\rho}(x)^{1/4}}{\sqrt{\rho(x)}} \sin \left[\frac{n \pi \tilde{\sigma}(x)}{\tilde{\sigma}(L)} \right] dx \ . \nonumber
\eeq

For $n \rightarrow \infty$ the leading contribution goes as
\beq
E_n^{(0)}+ E_n^{(1)}  \rightarrow  \frac{n^2 \pi^2}{\sigma(L)^2} \left[ \frac{\sigma(L)^2}{\tilde{\sigma}(L)^3} 
\int_{-L}^{+L} \frac{\tilde{\rho}(x)^{3/2}}{\rho(x)} dx \right] \ .
\eeq

Deviations of the term inside the parenthesis from $1$ imply violations of the correct leading asymptotic behaviour.

\section{Iterative methods}
\label{sec:iterative}

Another approach which can be used to obtain precise approximations to the solutions of the equation for an 
inhomogeneous string has been recently discussed in ref.~\cite{Amore10b}. In that paper we have proved three 
theorems which allow one to build iteratively the solutions of the fundamental and excited modes of the string 
starting from a trial solution, fulfilling the appropriate boundary conditions.

In particular, the main theorem, Theorem 1, may be view as a generalization of the well-known power method of linear 
algebra to the inverse operator of the inhomogeneous Helmholtz equation: the sequence of iterations produces 
approximate solutions which provide a sequence of increasingly tighter bounds for the energy of the fundamental mode
when the expectation value of the energy is calculated. The only limitation in the application of these theorems is
given by the ability of performing the relevant integrations explicitly.

We state now  the three theorems, which are proved in ref.~\cite{Amore10b}. Applications of these theorems will be 
considered in the following section.

\newtheorem{thm}{Theorem}
\begin{thm}
\label{theo1}
Let $\xi_0(x)$ be an arbitrary function defined on the interval $(-L,L)$ (being $\xi_0(x)$ arbitrary we assume that it
has nonzero overlap with the true fundamental solution). For $n\rightarrow \infty$ the sequence of functions 
\beq
\xi_n(x) \equiv \sqrt{\rho(x)} \int_{-L}^x dy \left[ \kappa_n - \int_{-L}^y dz \sqrt{\rho(z)} \xi_{n-1}(z) \right] \ ,
\label{iter}
\eeq
with $\kappa_n =  \frac{1}{2L} \int_{-L}^{L} dy \int_{-L}^{y} dz \sqrt{\rho(z)} \xi_{n-1}(z)$, converges to the fundamental 
mode of the Helmholtz equation (\ref{sec:dpt:1}) with Dirichlet boundary conditions, $\xi_n(\pm L) = 0$:
\beq
\Psi_0(x) = \lim_{n\rightarrow \infty} \tilde{\xi}_n(x) \ ,
\eeq
where $\tilde{\xi}_n(x) \equiv \frac{\xi_n(x)}{\sqrt{\rho(x)}}$.
\end{thm}

Using this theorem we may write the bounds
\beq
\frac{\int_{-L}^{+L} \xi_1(x) \xi_{0}(x)dx}{\int_{-L}^{+L} \xi_1^2(x) dx} \geq \dots \geq 
\frac{\int_{-L}^{+L} \xi_n(x) \xi_{n-1}(x)dx}{\int_{-L}^{+L} \xi_n^2(x) dx}   \geq  E_1 \ ,
\eeq
where $E_1$ is the exact energy of the fundamental mode of eq.~(\ref{sec:dpt:1}).

\begin{thm}
\label{theo2}
Let $\Xi^{(0)} \equiv \left\{ \xi_0^{(1)}(x), \dots, \xi_0^{(N)}(x)\right\}$ be a set of arbitrary functions defined on the interval $(-L,L)$. 
We assume that these functions have a nonzero overlap with the first $N$ solutions of the inhomogeneous Helmholtz equation.
For convenience we assume that the functions are ordered as
\beq
0 < \frac{\int_{-L}^{L}  \xi_{0}^{(1)}(x) \hat{O} \xi_{0}^{(1)}(x) dx }{\int_{-L}^{L}  \xi_{0}^{(1)}(x)^2 dx} \leq \dots
\leq \frac{\int_{-L}^{L}  \xi_{0}^{(N)}(x) \hat{O} \xi_{0}^{(N)}(x) dx }{\int_{-L}^{L}  \xi_{0}^{(N)}(x)^2 dx} \ .
\eeq

Consider the sequence of functions 
\beq
\xi^{(j)}_1(x) \equiv \sqrt{\rho(x)} \int_{-L}^x dy \left[ \kappa_1^{(j)} - \int_{-L}^y dz \sqrt{\rho(z)} \xi^{(j)}_{0}(z) \right] \ ,
\label{step1}
\eeq
with $\kappa_1^{(j)} =  \frac{1}{2L} \int_{-L}^{L} dy \int_{-L}^{y} dz \sqrt{\rho(z)} \xi_{0}^{(j)}(z)$ and $j=1,\dots, N$.

Let
\beq
\bar{\xi}_1^{(j)}(x) = \xi_1^{(j)}(x) - \sum_{k=1}^{j-1} 
\frac{\int_{-L}^{L} \bar{\xi}_1^{(k)}(x) {\xi}_1^{(j)}(x) dx}{\int_{-L}^{L} \bar{\xi}_1^{(k)}(x)^2 dx} \bar{\xi}_1^{(k)}(x)
\label{step2}
\eeq
and $\Xi^{(1)} \equiv \left\{ \bar{\xi}_1^{(1)}(x), \dots, \bar{\xi}_1^{(N)}(x)\right\}$. 
We call $\Xi^{(n)}$ the set of functions obtained after $n$ iterations. Then, for 
$n \rightarrow \infty$, $\Xi^{(n)}$ converges to the $N$ lowest eigenfunctions of eq.~(\ref{sec:dpt:1}).
\end{thm}

\begin{thm}
\label{theo3}
Let $\eta_0(x)$ be an arbitrary function fulfilling Dirichlet boundary conditions on $[-L,L']$, with $-L < L' < L$.
The sequence of functions
\beq
\eta_n(x) \equiv \sqrt{\rho(x)} \int_{-L}^x dy \left[ \tilde{\kappa}_n - \int_{-L}^y dz \sqrt{\rho(z)} \eta_{n-1}(z) \right] \ ,
\label{iter3}
\eeq
with $\tilde{\kappa}_n =  \frac{1}{L+L'_n} \int_{-L}^{L'_n} dy \int_{-L}^{y} dz \sqrt{\rho(z)} \eta_{n-1}(z)$, converges 
to a solution of equation (\ref{sec:dpt:1}) for $n\rightarrow \infty$ if $L'$ is chosen so that
$\eta_n(L)=0$.
\end{thm}

This theorem can be used to obtain estimates for the energies of the excited solutions of eq.~(\ref{sec:dpt:1}):
\beq
E_k \approx \frac{\int_{-L}^{+L} \eta^{(k)}_n(x) \eta^{(k)}_{n-1}(x)dx}{\int_{-L}^{+L} \left(\eta^{(k)}_n(x)\right)^2 dx}   
\eeq
where $E_k$ is the exact energy of $k^{th}$ state and $\eta_n^{(k)}(x)$ is the function obtained after $n$ iterations
and corresponding to the value of $L'$ obtained from the $k^{th}$ solution to the equation $\eta_n(L)=0$. Notice that
the $E_k$ for $k>0$ do not provide upper bounds to the exact energies.

\section{Some applications}
\label{sec:applications}

In this section we consider some applications of the approaches described in the previous sections.

\subsection{The ratio of the first two eigenvalues of an inhomogeneous string}

A first application concerns the calculation of the ratio of the first two eigenvalues of a string 
of variable density. Huang has proved in ref.~\cite{Huang99} that for strings with concave densities the ratio 
$E_2/E_1$ is minimized when the density is constant, whereas for strings with convex densities the ratio is 
maximized when the density is constant.

We may check what DPT and WKBPT can tell us concerning this problem. Working to first order and  assuming
$\rho(x) = 1+ \delta\rho(x)$ with $\int_{-L}^{+L}\delta\rho(x) dx = 0$, we obtain the approximation:
\beq
\frac{E_2}{E_1} \approx 4 \left[ 1 - \langle 2 | \delta\rho | 2 \rangle + \langle 1 | \delta\rho | 1\rangle \right] \ ,
\eeq
which holds for $|\delta\rho|\ll 1$.  

To recover the result of Huang we rewrite our formula as
\beq
\frac{E_2}{E_1} \approx 4 \left[ 1 - \langle 2 | \rho | 2 \rangle + \langle 1 | \rho | 1\rangle \right] \ .
\eeq

When the density takes larger values at the ends of the string one has that 
$\langle 2 | \rho | 2 \rangle - \langle 1 | \rho | 1\rangle > 0$, since the solution for the second mode takes 
larger values (in absolute value) in this region with respect to the solution of the first mode: in this case the 
ratio of the first two eigenvalues is maximized when the density is constant; conversely, when the density takes 
larger values at the center of the string, then $\langle 2 | \rho | 2 \rangle - \langle 1 | \rho | 1\rangle < 0$ 
and the opposite behaviour takes place.

Let us now consider the same problem using WKBPT: we work to first order and obtain the ratio
\beq
\frac{E_2}{E_1} \approx 4  + \frac{\sigma(L)^2}{\pi^2} \ \left( \langle 2 | V | 2\rangle  -4 \langle 1 | V | 1\rangle  \right) \ .
\eeq

As before we assume $\rho(x) = 1+ \delta\rho(x)$, with $\int_{-L}^{+L}\delta\rho(x) dx = 0$: 
assuming that $|\delta\rho| \ll 1$ we may approximate $V(x) \sigma'(x) \approx \frac{1}{4} \delta\rho''(x)$ 
and obtain a simple expression for the ratio: 
\beq
\frac{E_2}{E_1} \approx 4  - \frac{L}{\pi^2} \ \int_{-L}^{+L}  \left[4 \sin^2 \left(\frac{\pi (x+L)}{2L}\right)
-\sin^2 \left(\frac{\pi (x+L)}{L}\right)\right] {\delta\rho''(x)}  dx \ . \nonumber 
\eeq

The term in parenthesis is positive and peaked at the center of the string, therefore the sign of the correction  to
the ratio of the energies depends essentially on the sign of $\delta\rho''(x)$ in proximity of $x=0$, confirming the analysis done
using the DPT.

\subsection{An exactly solvable string}
\label{subsec:horgan}

Horgan and Chan give in ref.~\cite{Horgan99} explicit examples of strings with nonuniform density whose frequencies and solutions
can be calculated analytically. We will use one of these examples to test the accuracy of the asymptotic formulas for the energies 
obtained before.

The density of this string is
\beq
\rho(x) = \frac{(a+2)^2}{2 (a (a+2) (2 x+1)+2)} \ ,
\label{horgan_rho}
\eeq
where $a>-1$ is a real parameter. The string has unit length and it is centered in the origin.

Horgan and Chan have shown that the frequencies of this string are obtained from the zeros of the trascendental equation
\beq
J_1(\lambda) Y_1((a+1) \lambda)-Y_1(\lambda) J_1((a+1) \lambda) = 0
\eeq
as
\beq
\omega = a \lambda .
\eeq
Here $J$ and $Y$ are Bessel functions of first and second kind respectively.

It is therefore possible to calculate a large number of energies ($E = \omega^2$) for this string with high accuracy, and therefore 
estimate the asymptotic behavior of the energy by means of a simple fit with the known leading asymptotic behavior of the energy
\beq
E_n \approx A_1 n^2 + A_2 + A_3/n^2 + \dots
\label{fit}
\eeq

\begin{table}[tbp]
\caption{Coefficients of the fit of the energies of the string with density (\ref{horgan_rho}) with $a=1$. The 
third column contains extrapolations obtained from the values of the fit.}
\bigskip
\label{table:fit}
\begin{center}
\begin{tabular}{|c|c|c|c|}
\hline
 & Fit &  Extrapolation & $|{\rm Fit-Extrapolation}|$ \\
\hline
$A_1$ &  9.8696044010893586188 & $\pi^2$       & $1.5 \times 10^{-102}$  \\
$A_2$ &  0.3750000000000000000 & $\frac{3}{8}$ & $1.1 \times 10^{-93}$  \\
$A_3$ & -0.0326523345722377584 & $- \frac{165}{512\pi^2}$   & $3.5 \times 10^{-85}$ \\ 
$A_4$ &  0.0091705849049749141 & $\frac{73179}{81920 \pi ^4}$ & $6.8 \times 10^{-77}$ \\
$A_5$ & -0.0058099597602144426 & $- \frac{81997443}{14680064 \pi ^6}$ & $9.1 \times 10^{-69}$ \\
\hline
\end{tabular}
\end{center}
\bigskip\bigskip
\end{table}

We have chosen $a=1$ and we have calculated numerically the first $10000$ energies of this string with high precision (each 
zero has been obtained to $200$ digits). We have then used the energies from $5000$ to $10000$ to obtain a fit  as in eq.~(\ref{fit})
involving the coefficients $A_1, \dots, A_{14}$. The first $5$ coefficients obtained with this fit are reported in  the second
column of Table \ref{table:fit}.

In the third column of this table we report  the explicit expressions to which 
the numerical coefficients in the second column seem to converge: the calculation of the first two coefficients 
$A_1$ and $A_2$ using WKBPT is straightforward and confirms the corresponding results in the third column.

A more interesting test is provided by the calculation of $A_3$ using WKBPT; for this problem we have that
$\sigma(L) = 1$ and $V(\sigma) = \frac{3}{4 (1+\sigma)^2}$. As we have seen the coefficient $A_3$ 
receives contributions both from the first order and the second order term in the perturbative
expansion:
\beq
A_3 = A_3^{(1)} + A_3^{(2)} \ ,
\eeq
where $A_3^{(1)}$ ($A_3^{(2)}$) is the contribution from first (second) order WKBPT.

The evaluation of the first term is trivial:
\beq
A_3^{(1)} = - \frac{\sigma(L)}{4\pi^2} \left[ V^{(1)}(\sigma(L)) -V^{(1)}(0)\right] = - \frac{21}{64 \pi^2} \approx  -0.03324601338 \ .
\eeq
Notice that this term accounts for most of the value of $A_3$ reported in the table.

Let us now consider the second contribution; in this case
\beq
\mathcal{G}(\upsilon, \sigma_0) &=& - \frac{i}{2} \left[V(\sigma_0+ i \upsilon)-V(\sigma_0- i \upsilon)\right] \nonumber \\
&=& -\frac{3 (\sigma_0+1) \upsilon}{2 \left((\sigma_0+1)^2+\upsilon^2\right)^2} \ ,
\eeq
where in this case $\sigma_0$ can take the values $0$ and $\sigma(L)=1$.

The calculation of $\mathcal{F}$ can be performed analytically and one obtains~\footnote{Notice that
\beq
V^{(2j+1)}(\sigma_0) = - \frac{3}{2} \frac{(j+1) (2j+1)!}{(1+\sigma_0)^{2j+3}} \nonumber  
\eeq
and therefore the series defining $\mathcal{F}(\upsilon,\sigma_0)$ is divergent.}:
\beq
\mathcal{F}(\upsilon,\sigma_0) &=& \frac{3}{8\Omega v^2} \left[ 2 \Omega \ {\rm Ci}(\Omega) \ \sin (\Omega)-2 \Omega
 \  {\rm Si}(\Omega) \ \cos (\Omega)+ \pi  \Omega \ \cos (\Omega)-2 \right]
\eeq
where $\Omega \equiv \frac{1+\sigma_0}{\upsilon}$ and ${\rm Si}$ and ${\rm Ci}$ are the sine and cosine integrals.

Having obtained an explicit expression for $\mathcal{F}$ we can then evaluate the $A_3^{(2)}$ coefficient,
which has been cast in the form of single series as:
\beq
A_3^{(2)} &=& \frac{\sigma(L)^2}{\pi^4} \sum_{k=1}^\infty \frac{1}{2k^2} 
\left\{ \left[ \mathcal{F}\left( \frac{ \sigma(L)}{\pi k} , \sigma(L) \right) + 
\mathcal{F}\left( \frac{ \sigma(L)}{\pi k} , 0 \right)  \right]^2 \right. \nonumber \\
&-& \left.  
\mathcal{F}\left( \frac{ \sigma(L)}{2\pi k} , \sigma(L) \right) 
\mathcal{F}\left( \frac{ \sigma(L)}{2\pi k} , 0 \right) \right\}  = \sum_{k=1}^\infty \gamma_k \ ,
\eeq
where
\beq
\gamma_k &\equiv& \frac{9}{128 \pi ^2} \left(\left(\pi  (-1)^k k (\pi -2 {\rm Si}(k \pi ))-2 \pi  k {\rm Si}(2 k \pi )+\pi ^2
   k-3\right)^2 \right. \nonumber \\
&-& \left. 8 \left(-2 \pi  k {\rm Si}(2 k \pi )+\pi ^2 k-1\right) \left(-4 \pi  k {\rm Si}(4 k
   \pi )+2 \pi ^2 k-1\right)\right) \ .
\eeq

The series for $A_3^{(2)}$ may be approximated by a partial sum over the first $N$ terms in the sum: for instance
when we restrict the sum to the first $N=5000$ terms we have
\beq
A_3 \approx A_3^{(1)} + \sum_{k=1}^{5000} \gamma_k  \approx -0.032652334572241609068 \ ,
\eeq                                      
which agrees with the result in the table to 14 digits.

A much better estimate may be obtained taking into account the asymptotic behavior of the $\gamma_k$ for $k \rightarrow \infty$:
\beq
\gamma_k \approx \frac{711}{512 \pi^6 k^4}
\eeq
and therefore we may write
\beq
A_3^{(2)} = \sum_{k=1}^\infty \gamma_k = \frac{79}{5120 \pi^2} + \sum_{k=1}^\infty \left[\gamma_k - \frac{711}{512 \pi^6 k^4} \right] \ .
\eeq

If we limit the sum to the first 5000 terms as before we obtain a more precise estimate for $A_3$:
\beq
A_3 &\approx&  A_3^{(1)} + \frac{79}{5120 \pi^2} + \sum_{k=1}^{5000} \left[\gamma_k - \frac{711}{512 \pi^6 k^4} \right] \nonumber \\
&\approx& -0.032652334572237758375 \ ,
\eeq
where all the digits agree with the numbers obtained for the fit (fewer digits are displayed for the results in the table).

\subsection{Casimir energy of an inhomogeneous string}
\label{subsec:casimir}

The energies (frequencies) of a inhomogeneous string depend both on the density of the string itself and on the extension
of the string; when the density or the length of the string (or both) are changed, these energies change in a way 
which can be described in our WKBPT scheme. In particular we have shown that one can extract the asymptotic contributions
to the energies of the string from the perturbative corrections in the limit $n \gg 1$, with a rigorous procedure
which does not involve any approximation: therefore one can obtain arbitrarily accurate expressions for the energies of the 
string as functions of $n$ working to high orders in WKBPT.

We will now apply this result to the calculation of the Casimir energy of an inhomogeneous string. The problem of calculating
the Casimir energy of a piecewise uniform string has been originally considered  by Brevik and Nielsen in ref.~\cite{Brevik90} 
and by later studied in a sequence of works~\cite{Li91,Brevik94,Brevik96,Lambiase00}. 

Here we consider a string of arbitrary density with Dirichlet boundary conditions;
the Casimir energy of the string is defined in a standard way as~\cite{Lambiase00}
\beq
E_C = \frac{1}{2} \sum_{n=1}^\infty \left( \omega_n - \overline{\omega}_n \right) \ ,
\eeq
where $\omega_n$ are the frequencies the particular string considered and $\overline{\omega}_n$ are the 
frequencies of the string when its length is made infinity and the density constant.

Working with the energy to order $1/n^2$ we can write explicit expressions for the frequencies:
\beq
\omega_n &\approx& \frac{\pi}{\sigma(L)} n +  \frac{\sigma(L)}{2n\pi} \langle V \rangle + \frac{\sigma(L)^3}{8 n^3\pi^3}  
\left[ \frac{4 \pi^2 A_3}{\sigma(L)^2}  - \langle V \rangle^2\right] + O\left[1/n^5\right]\ ,
\eeq
where the general form of $A_3$ has been already derived using WKBPT.

Coming to the calculation of the Casimir energy, we may use a simple cutoff function $f(n) = e^{-\alpha n}$ as done in ref.~\cite{Brevik90} 
to regularize the divergent behaviour of the infinite series:
\beq
E_C &=& \frac{1}{2} \sum_{n=1} f(n) \left( \omega_n - \overline{\omega}_n \right) \nonumber \\
&=& \frac{1}{2} \left\{ \left(- \frac{1}{12} + \frac{1}{\alpha^2} + O\left[\alpha^2\right] \right) \frac{\pi}{\sigma(L)} 
+ \left(- \log\alpha + O\left[\alpha\right]\right)  \frac{\sigma(L)}{2\pi} \langle V \rangle \nonumber  \right. \\
&+& \left.   \zeta(3)  \frac{\sigma(L)^3}{8 \pi^3}  \left[ \frac{4 \pi^2 A_3}{\sigma(L)^2}  - \langle V \rangle^2\right] 
+\dots \right\} - \frac{1}{2} \sum_{n=1}^\infty \overline{\omega}_n  \ .
\eeq

Although it is not wise to attach a physical meaning to this expression until a consistent renormalization is carried out 
(which is beyond the goals of the present discussion), leading to a result which is free from divergencies, we may observe 
that the terms stemming from the higher order expansion in powers of $1/n$ lead to unambigous finite expressions which are
functionals of the density.  This is the case of the term proportional to $\zeta(3)$. The divergent behavior (in the limit 
$\alpha \rightarrow 0$) is contained in the first two terms and it should be cancelled by the last term.

\subsection{An asymmetric density}
\label{subsec:bounds}

We now study the example considered by Bender and Orszag at pag. 491 of ref.~\cite{BO78}, corresponding to the density 
(we center the string in the origin and set $L=\pi/2$)
\beq
\rho(x) = \left( x + \frac{3}{2}\pi \right)^4 \ .
\label{string_dens}
\eeq

In this case
\beq
\sigma(x) = \frac{8 x^3+36 \pi  x^2+54 \pi ^2 x+19 \pi ^3}{24}
\eeq
and
\beq
V(x) = -\frac{2}{\left(x+\frac{3 \pi }{2}\right)^6}  = -\frac{2}{\left(3 \sigma +\pi ^3\right)^2} \ .
\eeq

Lindblom and Robiscoe have also used this example in ref.~\cite{Lindblom91} to test the accuracy of their method.

Before discussing in detail the application of the results of the previous sections to this model, we should obtain accurate 
approximations to the solutions of this inhomogeneous string, which will be needed in order to assess the usefulness of our methods.

A general approach, which can be used for strings of arbitrary density, is based on the discretization of the
Helmholtz equation on a homogeneous grid using a collocation approach based on Little Sinc Functions (see for instance
ref.~\cite{Amore10b}). Using this approach one is able to obtain very precise values for the energies and eigensolutions of a large 
number of states of the inhomogeneous string: in particular, the energy of the fundamental mode of the string with the density 
of eq.~(\ref{string_dens}) using a fine grid of $2500$ points is
\beq
E_0^{LSF_{2500}} = 0.001744013543 \ ,
\eeq
where all the digits displayed are expected to be exact. Notice that this approach is more precise for the lowest states.

While the collocation approach is general, for the example of a string with density (\ref{string_dens}) we may obtain 
extremely precise results observing that the matrix elements of the operator $\hat{O} \equiv \frac{1}{\sqrt{\rho(x)}}
(-\frac{d^2}{dx^2}) \frac{1}{\sqrt{\rho(x)}}$ in the basis of the eigenfunctions of $\hat{Q}$ can be calculated analytically. 
This procedure amounts to using a {\sl spectral } method compared with the {\sl pseudospectral } method used in collocation.

We report here the results:
\beq
\langle n | \hat{O} | m \rangle &=& \delta_{nm} \left[\frac{9 n^2}{49 \pi^4} 
-\frac{4 n {\rm Ci}\left(\frac{2 n \pi }{7}\right) \sin \left(\frac{2 \pi n}{7}\right)}{49 \pi ^5}
+\frac{4 n {\rm Ci}\left(\frac{16 n \pi }{7}\right) \sin \left(\frac{2 \pi  n}{7}\right)}{49 \pi ^5} \right. \nonumber \\
&+& \left. \frac{4 n {\rm Si}\left(\frac{2 n \pi }{7}\right) \cos \left(\frac{2\pi  n}{7}\right)}{49 \pi ^5}
-\frac{4 n {\rm Si}\left(\frac{16 n \pi}{7}\right) \cos \left(\frac{2 \pi  n}{7}\right)}{49 \pi ^5}\right] \nonumber
\eeq
for $n = m$ and
\beq
\langle n | \hat{O} | m \rangle  &=& \frac{2}{49 \pi ^5} 
\left[ \left((m-n) \left({\rm Ci}\left(\frac{(m-n) \pi}{7} \right)
       -{\rm Ci}\left(\frac{8 (m-n) \pi}{7} \right)\right) \sin \left(\frac{\pi  (m-n)}{7}\right)  \nonumber \right. \right.\\
&+& \left. \left. (m+n) \left({\rm Ci}\left(\frac{8 (m+n) \pi}{7} \right)-{\rm Ci}\left(\frac{(m+n) \pi}{7}\right)\right) 
\sin \left(\frac{\pi  (m+n)}{7}\right) \nonumber \right. \right.\\
&+& \left. \left. (m-n) \left({\rm Si}\left(\frac{8 (m-n) \pi}{7} \right)-{\rm Si}\left(\frac{(m-n) \pi}{7} \right)\right) 
\cos \left(\frac{\pi (m-n)}{7} \right) \nonumber \right. \right.\\
&+& \left. \left.(m+n) \left({\rm Si}\left(\frac{(m+n) \pi}{7}\right)-{\rm Si}\left(\frac{8 (m+n) \pi}{7} \right)\right) 
\cos\left(\frac{\pi  (m+n)}{7}\right)\right)\right] \nonumber \ ,
\eeq
for $n\neq m$. Here $n, m = 1, 2, \dots$ and ${\rm Ci}(x)$, ${\rm Si}(x)$ are the cosine and sine integrals
respectively.

\begin{figure}
\begin{center}
\bigskip\bigskip\bigskip
\includegraphics[width=11cm]{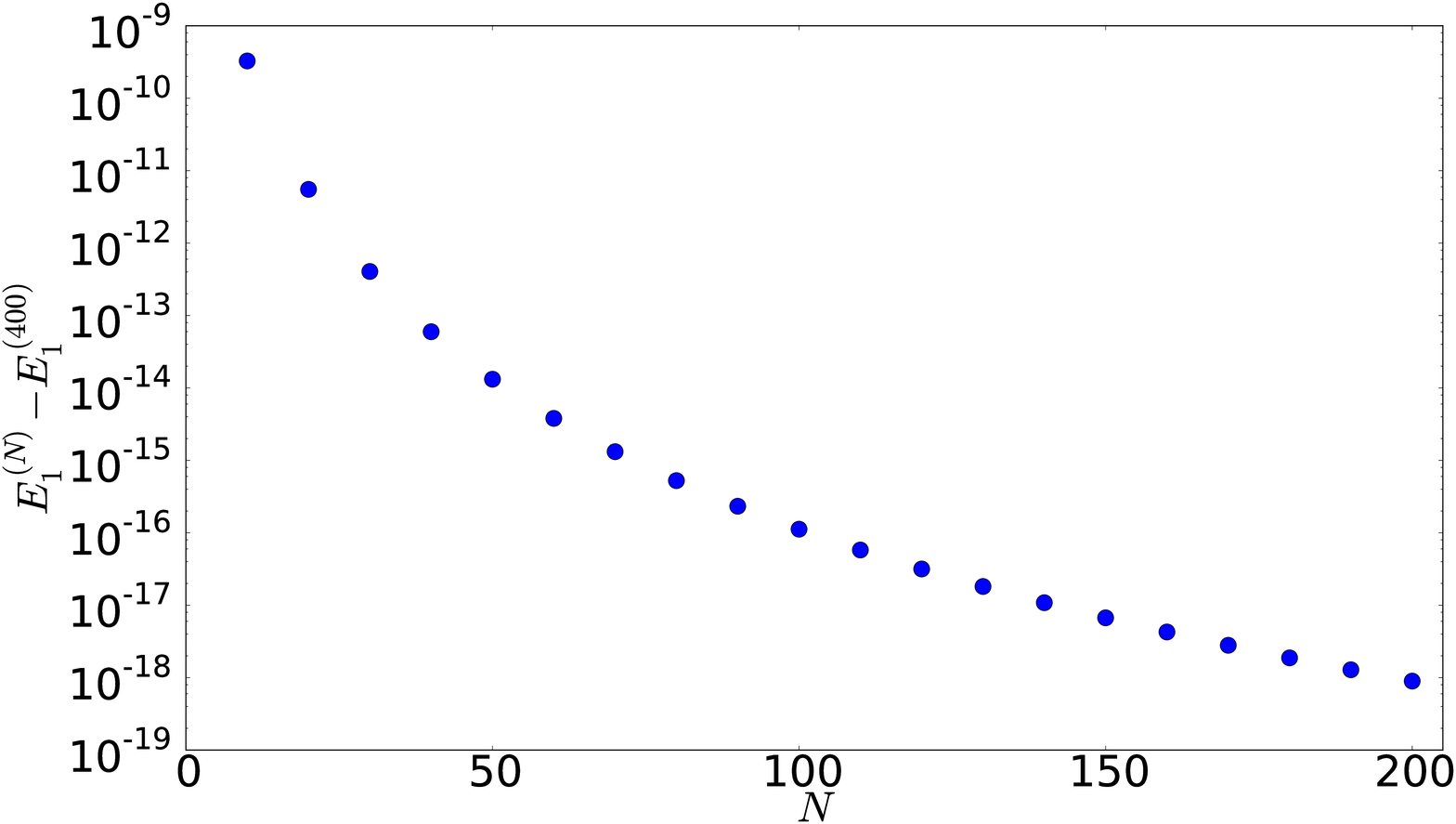}
\caption{The difference $E_1^{(N)}-E_1^{(400)}$ for the lowest eigenvalue of the matrices representing $\hat{O}$ on 
a $N$ dimensional subspace.}\bigskip
\label{Fig_0}
\end{center}
\end{figure}

Using these matrix elements we have obtained a $N \times N$ matrix representing $\hat{O}$  
in the subspace spanned by the lowest $N$ eigenfunctions of $\hat{Q}$; the eigenvalues and eigenvectors 
of this matrix are expected to converge to the exact eigenvalues and eigenfunctions of $\hat{O}$ as the 
dimension of the subspace is increased. This behavior is illustrated in Fig.~\ref{Fig_0} where we 
have calculated $E_1^{(N)}$ for matrices up to size $200 \times 200$ and we have then plotted the difference 
$E_1^{(N)}-E_1^{(400)}$ using the very precise numerical eigenvalue of the $400 \times 400$ matrix. This 
plot shows that the error for the lowest eigenvalue of the $200 \times 200$ matrix is of order $10^{-18}$,
and therefore much smaller than the one obtained with LSF.

In this case we have
\beq
E_1^{(200)} = 0.00174401354381855
\label{e1_200}
\eeq
where all the digits are expected to be correct.

\begin{figure}
\begin{center}
\bigskip\bigskip\bigskip
\includegraphics[width=11cm]{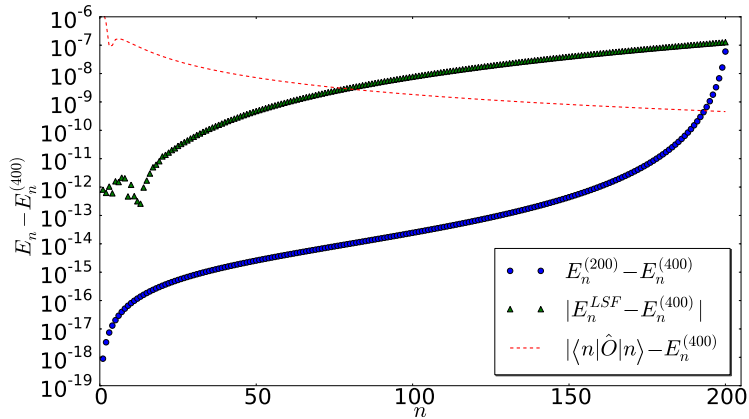}
\caption{The difference $E_n-E_n^{(400)}$ for the lowest $200$ eigenvalues of the string of density (\ref{string_dens}).
The circles correspond to $E_n$ calculated taking the eigenvalues of the $200 \times 200$ matrix representing $\hat{O}$, whereas the
triangles correspond to $E_n$ obtained using LSF with a grid of $2500$ points. The dashed curve corresponds to the difference
$| \langle n | \hat{O}| n \rangle - E_n^{(400)}|$.}
\label{Fig_0b}
\end{center}
\end{figure}

This approach allows one to obtain very precise approximations also for the excited states of the string: in Fig.~\ref{Fig_0b}
we see that the 200 eigenvalues of the $200 \times 200$ matrix representing $\hat{O}$ (circles) are very precise, even for the highest 
states. The few highest eigenvalues in the plot are obtained even more precisely using the diagonal matrix elements 
$\langle n | \hat{O} | n \rangle$ (dashed line), corresponding to using WKBPT to first order.

Actually one can use this spectral approach to obtain accurate results also for arbitrarily excited states: the 
crucial observation is that the matrix representing $\hat{O}$ becomes diagonally dominated for highly excited states.
This can be understood in general terms remembering that the functions $\phi_n(x)$ of eq.(\ref{sec:wkbpt:1}) become 
eigenfunctions of $\hat{O}$ for $n \rightarrow \infty$. 

Therefore, if we build a square $(2 \bar{n} +1) \times (2\bar{n}+1)$ matrix, centered on the $n$ state, its eigenvalues and
eigenvectors will provide increasingly better approximations to the exact eigenvalues and eigenfunctions of $\hat{O}$ as 
$n$ goes to infinity. In this way one may obtain precise estimates of the asymptotic behavior of
the energies of this problem for $n \rightarrow \infty$.

For example, we may consider the three states corresponding to $n=10^{5}-1$, $10^{5}$ and $10^{5}+1$ and build a matrix 
centered at each of these states with $\bar{n} = 100$. 

To calculate the eigenvalues we have worked with the numerical matrices with a precision of $100$ digits and we have then extracted
the coefficients of the asymptotic expansion with a fit, obtaining 
\beq
E_n &\approx& 0.0018855885773910003409 \ n^2 - 0.00026004036832396307366 \nonumber \\
&+& \frac{0.00075459566769573983984}{n^2} + \dots \ .
\eeq

These numerical coefficients should be compared with the exact coefficients calculated with WKBPT:
\beq
E_n = A_1 n^2 + A_2 + \frac{A_3}{n^2} + \dots 
\eeq
where
\beq
A_1 &=& \frac{9}{49 \pi ^4} \approx 0.0018855885773910003409 \\
A_2 &=& -\frac{1}{4 \pi ^6} \approx -0.00026004036832396307402 \ , 
\eeq
and
\beq
A_3 &=& A_3^{(1)} + A_3^{(2)} \ .
\eeq

The contribution to $A_3$ stemming from first order is easily calculated and reads:
\beq
A_3^{(1)} &=&  - \frac{\sigma(L)}{4\pi^2} \left[ V^{(1)}(\sigma(L)) -V^{(1)}(0)\right] = \frac{3577}{512 \pi^8} \approx 0.000736 \ .
\eeq

As we have seen before the contribution to $A_3$ stemming from second order involves a series
\beq
A_3^{(2)} &=& \frac{\sigma(L)^2}{\pi^4} \sum_{k=1}^\infty \frac{1}{2k^2} 
\left\{ \left[ \mathcal{F}\left( \frac{ \sigma(L)}{\pi k} , \sigma(L) \right) + 
\mathcal{F}\left( \frac{ \sigma(L)}{\pi k} , 0 \right)  \right]^2 \right. \nonumber \\
&-& \left.  
\mathcal{F}\left( \frac{ \sigma(L)}{2\pi k} , \sigma(L) \right) 
\mathcal{F}\left( \frac{ \sigma(L)}{2\pi k} , 0 \right) \right\}  \equiv \sum_{k=1}^\infty \gamma_k \ ,  
\eeq
and requires the calculation of $\mathcal{F}$.

For this example we find
\beq
\mathcal{G}(\upsilon,\sigma_0) &=& \frac{12 \left(3 \sigma_0+\pi^3\right) \upsilon}{\left(9 \left(\sigma_0^2+\upsilon^2\right)+6 \pi^3
   \sigma_0+\pi^6\right)^2}
\eeq
and
\beq
\mathcal{F}(\upsilon,\sigma_0)  &=& \frac{-2 \Lambda {\rm Ci}(\Lambda) \sin (\Lambda)+2 \Lambda {\rm Si}(\Lambda) 
\cos (\Lambda)-\pi  \Lambda \cos (\Lambda)+2}{9 \upsilon^2 \Lambda}
\eeq
where 
\beq
\Lambda \equiv \frac{\pi^3+3 \sigma_0}{3\upsilon} \ .
\eeq

After substituting the expression for $\mathcal{F}$ inside the expression for $A_3^{(2)}$ we obtain
an explicit expression for $\gamma_k$:
\beq
\gamma_k &\equiv& \frac{1}{14112 \pi ^8} \left[
\left(8 \pi  k {\rm Ci}\left(\frac{k \pi }{7}\right) \sin \left(\frac{\pi  k}{7}\right)
     +8 \pi  k {\rm Ci}\left(\frac{8 k \pi }{7}\right) \sin \left(\frac{8 \pi  k}{7}\right) \right. \right.\nonumber \\
 &-& \left.\left. 8 \pi  k {\rm Si}\left(\frac{k \pi }{7}\right) \cos \left(\frac{\pi  k}{7}\right)
  -8 \pi  k {\rm Si}\left(\frac{8 k \pi }{7}\right) \cos \left(\frac{8 \pi  k}{7}\right) \right. \right.\nonumber \\
 &+& \left. \left. 4 \pi ^2 k \cos\left(\frac{\pi  k}{7}\right)+4 \pi ^2 k \cos \left(\frac{8 \pi  k}{7}\right)-63\right)^2 
\right. \nonumber \\
&-& \left. 32 \left(2 \pi  k {\rm Ci}\left(\frac{2 k \pi }{7}\right) \sin \left(\frac{2 \pi  k}{7}\right)-2 \pi
    k {\rm Si}\left(\frac{2 k \pi }{7}\right) \cos \left(\frac{2 \pi  k}{7}\right) \right. \right. \nonumber \\
&+& \left.\left.\pi ^2 k \cos\left(\frac{2 \pi  k}{7}\right)-7\right)  \cdot  \left(16 \pi  k {\rm Ci}\left(\frac{16 k \pi }{7}\right)
   \sin \left(\frac{16 \pi  k}{7}\right) \right. \right. \nonumber \\
&-& \left.\left. 16 \pi  k {\rm Si}\left(\frac{16 k \pi }{7}\right) 
\cos\left(\frac{16 \pi  k}{7}\right)+8 \pi ^2 k \cos \left(\frac{16 \pi  k}{7}\right)-7\right) \right]  \nonumber \ .
\eeq

We may estimate $A_3$ by restricting the series in $A_3^{(2)}$ to the first $N$ terms. For example, using $N=8000$ we obtain
\beq
A_3 = A_3^{(1)}+ A_3^{(2)} \approx  \frac{3577}{512 \pi^8}  + \sum_{k=1}^{8000} \gamma_k \approx 0.00075459642403370616977 \ .
\eeq

A much better estimate may be obtained taking into account the asymptotic behavior of the $\gamma_k$ for $k \rightarrow \infty$:
\beq
\gamma_k \approx \frac{631561441}{294912 \pi ^{12} k^4}
\eeq
and therefore we may write
\beq
A_3^{(2)} = \sum_{k=1}^\infty \gamma_k = \frac{631561441}{26542080 \pi ^8} + \sum_{k=1}^\infty \left[\gamma_k - 
\frac{631561441}{294912 \pi ^{12} k^4} \right] \ .
\eeq

If we limit the sum to the first 8000 terms as before we obtain a more precise estimate for $A_3$:
\beq
A_3 &\approx&  A_3^{(1)} + \frac{631561441}{26542080 \pi ^8} + \sum_{k=1}^{8000} \left[\gamma_k - 
\frac{631561441}{294912 \pi ^{12} k^4} \right] \nonumber \\
&\approx& 0.00075459642403521434583 \ .
\eeq

Therefore we see that the coefficients obtained with the fit are in very good agreement with the theoretical
values obtained with WKBPT.

We will now use the very precise numerical results at our disposal to discuss the WKB-perturbation method 
and its improved version, which we have described in the previous sections.

We first concentrate on the bounds for the energy of the fundamental mode, starting with the one obtained using
DPT, which largely overestimates the exact result
\beq
E_1 \leq \frac{4}{15 \pi} \ {\rm Si}(2 \pi )-\frac{4}{15 \pi} \  {\rm Si}(4 \pi ) +\frac{7}{30 \pi^4}+\frac{1}{15 \pi^2} 
\approx 0.002868007277 \ .
\eeq

We should not be surprised by this result since the density of the string varies strongly on the domain and therefore 
density perturbation theory is not applicable.

We now discuss the first bound obtained WKBPT. First we make an important observation: since $V(\sigma)$ is negative on the 
whole string, then $E_n^{(1)}$ in eq.~(\ref{bound_wkbpt1}) is also negative. Moreover, for the modes of the string
which are arbitrarily excited the ratio $E_n^{(1)}/E_n^{(0)}$ becomes arbitrarily small, while $E_n^{(0)}$ becomes arbitrarily 
accurate as $n \rightarrow \infty$, thus we may conclude that the $E_n^{(0)}$ are upper bounds to the energies of the string
as $n \rightarrow \infty$. We have numerically tested this conjecture using the eigenvalues of the $400 \times 400$ matrix 
representing $\hat{O}$ and we have found that it actually holds even for the intermediate states.

The WKBPT bound is readily obtained using the matrix element previously calculated:
\beq
E_1 &\leq& \frac{9 \pi }{49 \pi ^5} + \frac{4}{49 \pi ^5} \left[ \left({\rm Ci}\left(\frac{16 \pi }{7}\right)-{\rm Ci}
\left(\frac{2 \pi }{7}\right)\right) \cos \left(\frac{3\pi }{14}\right) \right. \nonumber \\
&+& \left. \left({\rm Si}\left(\frac{2 \pi }{7}\right)-{\rm Si}\left(\frac{16 \pi }{7}\right)\right) 
\sin\left(\frac{3 \pi }{14}\right) \right] \approx  0.001745807766 \ ,
\eeq
which is much tighter than the bound obtained with DPT.
It is also interesting to compare this result with the WKB result to first order:
\beq
E_1^{{\rm WKB} (0)} +E_1^{{\rm WKB} (1)} = \frac{9}{49 \pi ^4}-\frac{1}{4 \pi ^6} \approx 0.00162555 \ ,
\eeq
which is less precise than the previous one and falls below the exact value.

As we have discussed before, one may still obtain better bounds using the first order WKBPT wave function as trial
solution: in the present example, the matrix elements $\langle k | V | l\rangle$ are known explicitly and therefore
the bound can be calculated easily. Notice that the variational bound holds also when a finite number of terms is 
used in the series appearing in $\langle \hat{O}\rangle_\Psi$. 

In Table \ref{table:1} we report the variational bounds obtained restricting the sum up to $k=N$ terms and compare the 
results with the "exact" numerical value quoted by Bender and Orszag and with the even more precise values that have been 
obtained using a collocation approach based on the LSF functions of ref.~\cite{Amore07} with grids with $2000$ and $2500$ 
points (the use of collocation in the numerical solution of the problem of an inhomogeneous string is illustrated 
in ref.~\cite{Amore10b}). Notice that for $N > 20$ the variational bound does not improve, signaling the need for a 
more refined ansatz.

\begin{table}[tbp]
\caption{Variational bounds using first order WKBPT to the energy of the fundamental mode of a string with 
density $\rho(x) = \left( x + \frac{3}{2}\pi \right)^4$. The notation $[-n]$ means $10^{-n}$.}
\bigskip
\label{table:1}
\begin{center}
\begin{tabular}{|c|c|c|}
\hline
$N$ & $E_{1}(N)$ & $E_1(N)-E_1^{\rm LSF_{2500}}$ \\
\hline
2 & 0.0017442945174961415 & 2.81 [-7]\\
3 & 0.0017440904942458490 & 7.70 [-8]\\
4 & 0.0017440396134339554 & 2.61 [-8]\\
5 & 0.0017440245258144354 & 1.10 [-8]\\
6 & 0.0017440189344824756 & 5.39 [-9] \\
7 & 0.0017440167007034352 & 3.16 [-9]\\
8 & 0.0017440156851693774 & 2.14 [-9]\\
9 &  0.0017440152042467934 & 1.66 [-9]\\
10 & 0.0017440149552016950 & 1.41 [-9]\\
20 & 0.0017440146360548171 & 1.09 [-9]\\
30 & 0.0017440146309786759 & 1.09 [-9]\\
40 & 0.0017440146306353234 & 1.09 [-9]\\
50 & 0.0017440146305891107 & 1.09 [-9]\\
\hline
ref.~\cite{BO78}  & 0.00174401 & \\
\hline
LSF$_{2000}$ & 0.0017440135432079554 &  \\
LSF$_{2500}$ & 0.0017440135430055502 & \\
\hline
\end{tabular}
\end{center}
\bigskip\bigskip
\end{table}

In Table \ref{table:2} we report the energies of selected modes of the string  considered by Bender and Orszag,
calculated with WKBPT to first order and second orders (second and fourth columns), with the asymtotic formulas  $A_1 n^2 + A_2$ and 
$A_1 n^2 + A_2 + \frac{A_3}{n^2}$  (third and fifth columns) and with the LSF collocation method with a grid with $2500$ points (sixth column). 
Clearly the second order WKBPT results are very precise even for the lowest modes: this fact signals that the dominant contributions to the
asymptotic coefficients $A_j$, with $j=4,5,\dots$, may be (at least in principle) calculated using the expression of the energy to second 
order in WKBPT, in analogy with what we have done in the calculation of $A_3$.

\begin{table}[tbp]
\caption{Comparison of the energies for selected modes of a string with density $\rho(x) = \left( x + \frac{3}{2}\pi \right)^4$
calculated with WKBPT to first and second orders (second and fourth columns), with the asymtotic formulas  $A_1 n^2 + A_2$ and 
$A_1 n^2 + A_2 + \frac{A_3}{n^2}$  (third and fifth columns) and with the LSF collocation method with a grid with $2500$ points (sixth column). 
The series in the second order WKBPT term is restricted to the 20 closest states.}
\bigskip
\label{table:2}
\begin{center}
\begin{tabular}{|c|c|c|c|c|c|}
\hline
$n$ & $E_{n}^{({\rm I})}$  & $A_1 n^2 + A_2$ & $E_{n}^{({\rm II})}$ & $A_1 n^2 + A_2 + \frac{A_3}{n^2}$ &  $E_n^{\rm LSF_{2500}}$ \\
\hline
1  &   0.00174581  &   0.00162555  &    0.00174405 &  0.00238014 &  0.00174401 \\
2  &   0.00734927  &   0.00728231  &    0.00734864 &  0.00747096 &  0.00734866  \\
3  &   0.01675247  &   0.01671026  &    0.01675237 &  0.01679410 &  0.01675238  \\
4  &   0.02993818  &   0.02990938  &    0.02993827 &  0.02995654 &  0.02993828   \\
5  &   0.04690044  &   0.04687967  &    0.04690060 &  0.04690986 &  0.04690060  \\
6  &   0.06763676  &   0.06762115  &    0.06763693 &  0.06764211 &  0.06763693  \\
7  &   0.09214593  &   0.09213380  &    0.09214609 &  0.09214920 &  0.09214609  \\
8  &   0.12042730  &   0.12041763  &    0.12042744 &  0.12042942 &  0.12042744  \\
9  &   0.15248051  &   0.15247263  &    0.15248064 &  0.15248195 &  0.15248064  \\
10  &  0.18830534  &   0.18829882  &    0.18830546 &  0.18830636 &  0.18830546  \\
20  &  0.75397717  &   0.75397539  &    0.75397721 &  0.75397728 &  0.75397721  \\
30  &  1.69677048  &   1.69676968  &    1.69677050 &  1.69677052 &  1.69677050  \\
40  &  3.01668214  &   3.01668168  &    3.01668215 &  3.01668216 &  3.01668215  \\
50  &  4.71371167  &   4.71371140  &    4.71371170 &  4.71371170 &  4.71371170  \\
\hline
\end{tabular}
\end{center}
\bigskip\bigskip
\end{table}

We will now discuss the bound obtained using the improved WKBPT approach (iWKBPT), eq. (\ref{bound_iwkbpt}).
We consider the trial density
\beq
\tilde{\rho}(x) =  \rho(x) \ \left[\sum_{j=0}^N a_j x^j\right]^2 \ ,
\label{trialdens}
\eeq
where the $a_j$ are $N+1$ real parameters to be determined variationally. 

Clearly choosing $a_0=1$  and $a_j = 0$ for $j>0$ (corresponding to $N=0$), $\tilde{\rho}(x)$ reduces to 
the physical density: in this case we recover the simpler bound of WKBPT. 

In Table \ref{table:4} we report the energy corresponding 
to a set of values of the "optimal" $a_i$ and the numerical value of the quantity 
$\frac{\sigma(L)^2}{\tilde{\sigma}(L)^3}\int_{-L}^{+L} \frac{\tilde{\rho}(x)^{3/2}}{\rho(x)} dx$, which determines the asymptotic 
behavior of the energies.

For $N=10$ the first 17 digits of the variational energy agree with those of the energy
of eq.~(\ref{e1_200}); the case $N=0$ corresponds to normal WKBPT, which uses 
the physical density as $\tilde{\rho}(x)$.

\begin{table}[tbp]
\caption{The variational estimate for the energy of the fundamental mode of the string of density 
$\rho(x) = \left( x + \frac{3}{2}\pi \right)^4$ and the quantity $\frac{\sigma(L)^2}{\tilde{\sigma}(L)^3}\int_{-L}^{+L} 
\frac{\tilde{\rho}(x)^{3/2}}{\rho(x)} dx$ determining the asymptotic behavior  of the spectrum. }
\bigskip
\label{table:4}
\begin{center}
\begin{tabular}{|c|c|c|c|c|c|c|}
\hline
$N$ &    $E_1^{\rm iWKBPT}$ &   $E_1^{\rm iWKBPT} - E_1^{(200)}$ & 
$\frac{\sigma(L)^2}{\tilde{\sigma}(L)^3}\int_{-L}^{+L} \frac{\tilde{\rho}(x)^{3/2}}{\rho(x)} dx$ \\
\hline
0   & 0.00174580776578874  & 1.79 [-6]  & 1.00000 \\
1   & 0.00174408029127495  & 6.67 [-8]  & 1.00076 \\
2   & 0.00174402710509883  & 1.36 [-8]  & 1.00088 \\
3   & 0.00174401553013316  & 1.99 [-9]  & 1.00088 \\
4   & 0.00174401377197412  & 2.28 [-10] & 1.00088 \\
5   & 0.00174401354435309  & 5.35 [-13] & 1.00088 \\
6   & 0.00174401354405299  & 2.34 [-13] & 1.00088 \\
7   & 0.00174401354384465  & 2.61 [-14] & 1.00088 \\
8   & 0.00174401354382102  & 2.47 [-15] & 1.00088 \\
9   & 0.00174401354381886  & 3.12 [-16] & 1.00088 \\
10  & 0.00174401354381859  & 3.70 [-17] & 1.00088 \\
\hline      
$E_1^{(200)}$ & 0.00174401354381855 &  & \\
\hline
\end{tabular}
\end{center}
\bigskip\bigskip
\end{table}

\begin{figure}
\begin{center}
\bigskip\bigskip\bigskip
\includegraphics[width=11cm]{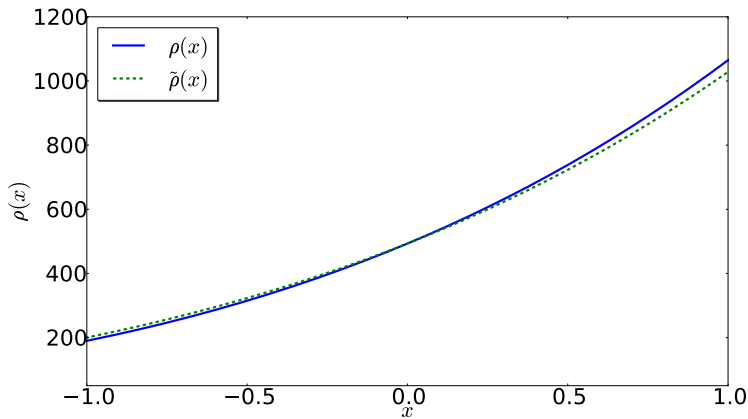}
\caption{Comparison between the physical density $\rho(x) = \left( x + \frac{3}{2}\pi \right)^4$ (solid line)
and the optimal trial density obtained with the iWKBPT bound with $N=10$ (dashed line).}
\label{Fig_1}
\end{center}
\end{figure}

In Fig.~\ref{Fig_1} we plot the physical density  $\rho(x) = \left( x + \frac{3}{2}\pi \right)^4$ (solid line)
and compare it with the optimal trial density obtained with the iWKBPT bound using $N=10$ parameters (dashed line).
This density approximates the effective density for the fundamental mode of $\hat{O}$.

Using the transformation discussed by in ref.~\cite{Gottlieb02} we may obtain refined bounds for the energy of the
fundamental mode of our string, studying an isospectral string of density
\beq
\bar{\rho}(x) = \frac{16 (\pi  \alpha +1)^2 (\pi  (2 \alpha  (2 x+\pi )+3)+2 x)^4}{(\alpha (2 x+\pi )+2)^8} \ . 
\eeq

In Table \ref{table:5} we report the values of these bounds obtained by minimizing the expectation value of $\bar{\hat{O}}$
in the fundamental mode of $\tilde{Q}$ where the density $\tilde{\rho}(x)$ has the same form considered before.
The minimization is done with respect to $\alpha$ and to the parameters in $\tilde{\rho}(x)$. In the fifth column 
we report the optimal value of $\alpha$, for a set of $N$ parameters $a_i$. 
The minimization of the expectation value for a set of $N+1$ parameters (the $N$ parameters $a_i$ and $\alpha$) 
is done using the $N$ parameters of the previous solution as a starting point. 

Notice that the result obtained for $N=0$ is roughly an order of magnitude more precise than the corresponding
result in Table \ref{table:4} and that it retains the exact asymptotic behavior. 

On the other hand the results obtained for $N \geq 7$ have a larger error than the corresponding results  obtained in 
Table \ref{table:4}: this means that the solution that we have found corresponds to a local minimum of the 
expectation value. On the other hand, we should observe that the asymptotic behavior of the spectrum obtained
in table \ref{table:5} for $N \geq 7$ more accurate than the corresponding behavior in Table \ref{table:4}.

\begin{figure}
\begin{center}
\bigskip\bigskip\bigskip
\includegraphics[width=11cm]{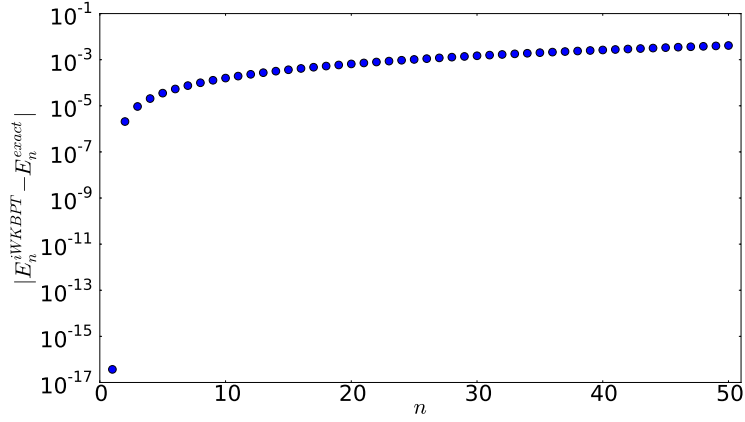}
\caption{$| E_n^{iWKBPT} - E_n^{exact} |$ for the case of a string with density $\rho(x) = \left( x + \frac{3}{2}\pi \right)^4$. The
iWKBPT results are obtained using an optimal density with $N=10$ parameters.}
\label{Fig_2}
\end{center}
\end{figure}

\begin{table}[tbp]
\caption{The variational estimate for the energy of the fundamental mode of the string of density 
$\bar{\rho}(x) = \frac{16 (\pi  \alpha +1)^2 (\pi  (2 \alpha  (2 x+\pi )+3)+2 x)^4}{(\alpha (2 x+\pi )+2)^8}$ 
and the quantity $\frac{\sigma(L)^2}{\tilde{\sigma}(L)^3}\int_{-L}^{+L} \frac{\tilde{\rho}(x)^{3/2}}{\rho(x)} dx$ determining 
the asymptotic behaviour of the spectrum. }
\bigskip
\label{table:5}
\begin{center}
\begin{tabular}{|c|c|c|c|c|c|c|}
\hline
$N$ &    $E_1^{\rm iWKBPT}$ &   $E_1^{\rm iWKBPT} - E_1^{(200)}$ & 
$\frac{\sigma(L)^2}{\tilde{\sigma}(L)^3}\int_{-L}^{+L} \frac{\tilde{\rho}(x)^{3/2}}{\rho(x)} dx$  & $\alpha$\\
\hline
0   & 0.00174418771063455 & 1.74 [-7]  & 1.00000 & -0.00840 \\
1   & 0.00174404317510904 & 2.96 [-8]  & 1.00415 &  0.01082 \\
2   & 0.00174401638359089 & 2.84 [-9]  & 1.00532 & -0.02892 \\
3   & 0.00174401358072024 & 3.69 [-11] & 1.00930 & -0.03528 \\
4   & 0.00174401356813542 & 2.43 [-11] & 1.02434 & -0.05076 \\
5   & 0.00174401354405135 & 2.33 [-13] & 1.00016 & -0.00565 \\
6   & 0.00174401354405132 & 2.33 [-13] & 1.00016 & -0.00565 \\
7   & 0.00174401354385037 & 3.18 [-14] & 1.00016 & -0.00566 \\
8   & 0.00174401354382192 & 3.37 [-15] & 1.00016 & -0.00566 \\
9   & 0.00174401354381896 & 4.10 [-16] & 1.00016 & -0.00566 \\
10  & 0.00174401354381860 & 5.05 [-17] & 1.00016 & -0.00566 \\
\hline      
$E_1^{(200)}$ & 0.00174401354381855 &  &  & \\
\hline
\end{tabular}
\end{center}
\bigskip\bigskip
\end{table}

\subsection{A string with rapidly oscillating density}

We will now study a string of unit length ($|x| \leq 1/2$) and with a rapidly oscillating density
\beq
\rho(x) = \left(2+ \sin (100 \pi  x)\right)^2 \ .
\eeq

In this case one has
\beq
\sigma(x) = 1+2 x+\frac{\sin ^2(50 \pi  x)}{50 \pi } \ .
\eeq

It is known that the solutions for strings with rapidly oscillating density have peculiar properties: 
in particular the solutions whose wavelength corresponds roughly to the typical size of the density 
oscillations are localized at the ends of the strings~\cite{CZ00b,ABR92,CZ00a}.

Before we apply the results of the previous sections to the study of this string we wish to obtain numerical
results which will be useful to assess the precision of the analytical methods: as for the previous
example we have used a collocation approach based on LSF with an homogeneous grid of $2500$ points. 
The grid spacing $h=2L/N$ is sufficiently fine to obtain precise results for the first few hundreds of 
states; in particular, the energy of the fundamental mode obtained with this grid is 
\beq
E_1^{LSF_{2500}} = 2.1931584082 \ .
\eeq
The quality of the variational bounds obtained with the different approaches will be established using 
this value.

\begin{figure}
\begin{center}
\bigskip\bigskip\bigskip
\includegraphics[width=11cm]{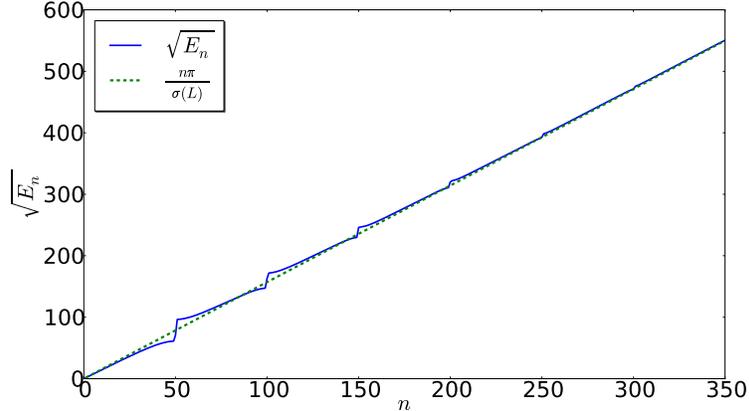}
\caption{Wave numbers of the string with density $\rho(x)=\left(2+ \sin (100 \pi  x)\right)^2$
obtained with LSF with a grid of $2500$ points. The dashed line is the asymptotic behavior
$k_n = n \pi/\sigma(L)$.}
\label{Fig_5}
\end{center}
\end{figure}

In Fig.~\ref{Fig_5} we display the numerical values of the square roots of the first $350$ energies of 
this string obtained with the LSF collocation using $2500$ points (plus symbols) and we compare them 
with the exact leading asymptotic behavior $k_n = n\pi/2$ (dashed line). Gaps are observed in correspondence 
to quantum numbers which are multiples of $50$: as we have anticipated these gaps occurr because
of the appearance of localized solutions at one of the ends of the string.

\begin{figure}
\begin{center}
\bigskip\bigskip\bigskip
\includegraphics[width=11cm]{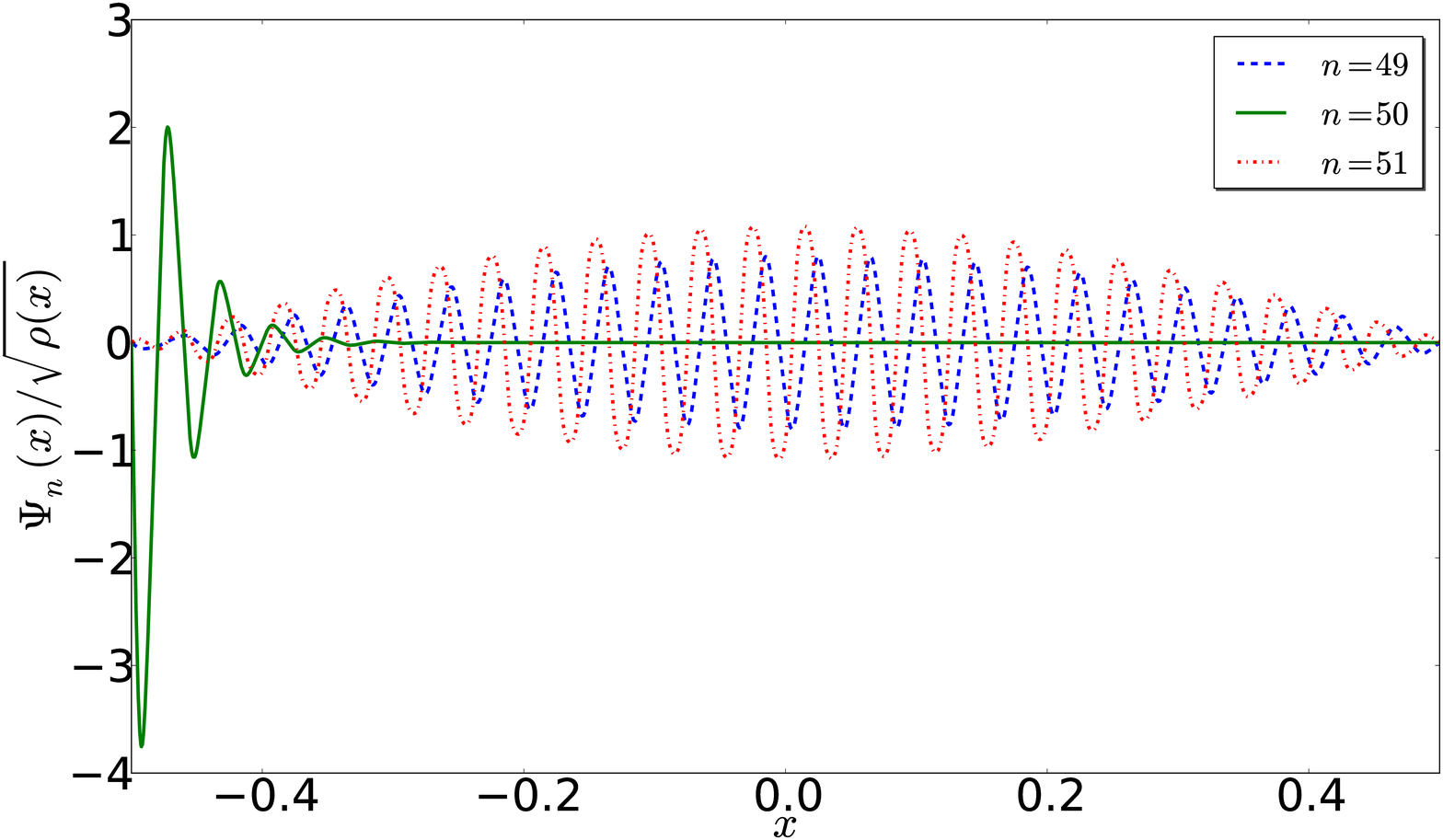}
\caption{Solutions for the modes $49$,$50$ and $51$ of the string with density $\rho(x)=\left(2+ \sin (100 \pi  x)\right)^2$
obtained with LSF with a grid of $2500$ points. }
\label{Fig_6}
\end{center}
\end{figure}

In Fig.~\ref{Fig_6} we display three solutions of the Helmholtz equation for the string with density 
$\rho(x)=\left(2+ \sin (100 \pi  x)\right)^2$, obtained using LSF collocation with a grid of $2500$ points.
The mode for $n=50$ is localized on the left end of the string, while the modes corresponding to
$n=49$ and $n=51$ extend over all the string. It is instructive to decompose these solutions in terms 
of the modes of an uniform string: we may expect that the modes of this string which are not localized may
be described in terms of few modes of an uniform string, whereas the modes of this string which are localized
necessarily involve a great number of modes of the uniform string.

This analysis can be carried out rapidly and efficiently using the results of the LSF collocation, and at no extra computational cost.
The fundamental observation is that the LSF are defined as
\beq
s_k(x) \equiv \frac{2L}{N} \sum_{n=1}^N \psi_n(x_k) \psi_n(x) \ ,
\eeq
where $\psi_n(x) \equiv \sqrt{\frac{1}{L}} \sin \frac{n\pi (x+L)}{2L}$ are the eigensolutions of the uniform string
on $-L \leq x \leq L$ with Dirichlet boundary conditions at $x = \pm L$. The $x_k \equiv 2L k/N$ are the uniformly
distributed grid points. 

A function $f(x)$ whose values are known at the grid points may then be approximated
at arbitrary points in the domain through the interpolation formula
\beq
f(x) &\approx&  \sum_{k=-N/2+1}^{N/2-1} f(x_k) s_k(x) \nonumber \\
&=& \sum_{n=1}^N \left[ \frac{2L}{N}  \sum_{k=-N/2+1}^{N/2-1} f(x_k) \psi_n(x_k) \right] \psi_n(x) \ ,
\eeq
where the terms inside the parenthesis are the approximate Fourier coefficients of $f(x)$.
In the case that we are considering $f(x_k)$ is just the $k^{th}$ entry of one of the eigenvectors of the $(N-1) \times (N-1)$
matrix that we have obtained from the discretization of the Helmholtz equation, multiplied by a factor $\sqrt{N/2L}$, and 
$\psi_n(x_k)$ is the $k^{th}$ entry of the vector obtained evaluating the $n^{th}$ eigensolution of the uniform string on the $k^{th}$ grid point.
The calculation of the Fourier coefficients thus involves (apart from a multiplicative factor) the multiplication of two vectors, 
which have been already calculated.

\begin{figure}
\begin{center}
\bigskip\bigskip\bigskip
\includegraphics[width=11cm]{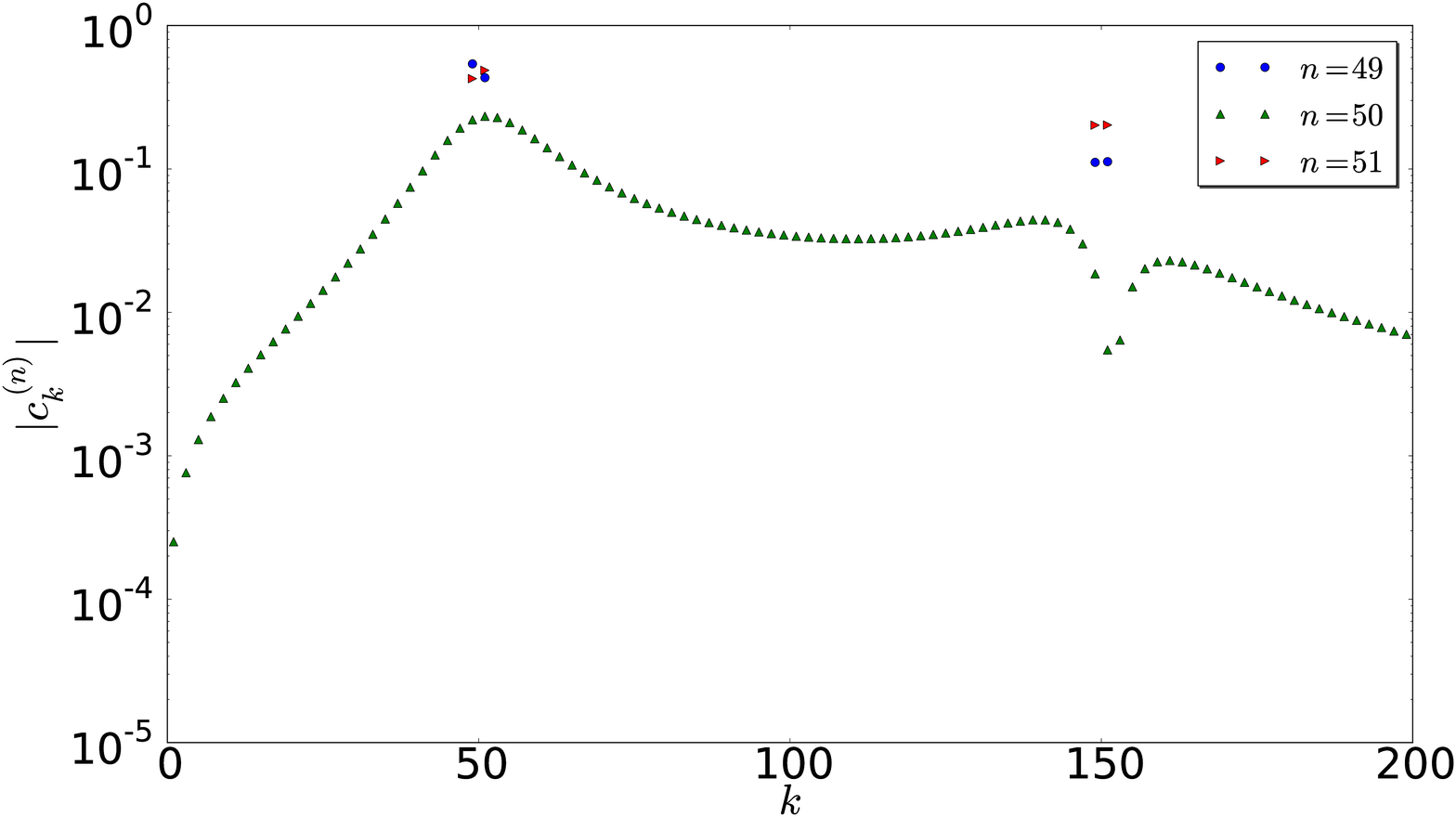}
\caption{Approximate odd Fourier coefficients for the modes $49$,$50$ and $51$ of the string with density 
$\rho(x)=\left(2+ \sin (100 \pi  x)\right)^2$ obtained with LSF with a grid of $2500$ points. }
\label{Fig_7odd}
\end{center}
\end{figure}

\begin{figure}
\begin{center}
\bigskip\bigskip\bigskip
\includegraphics[width=11cm]{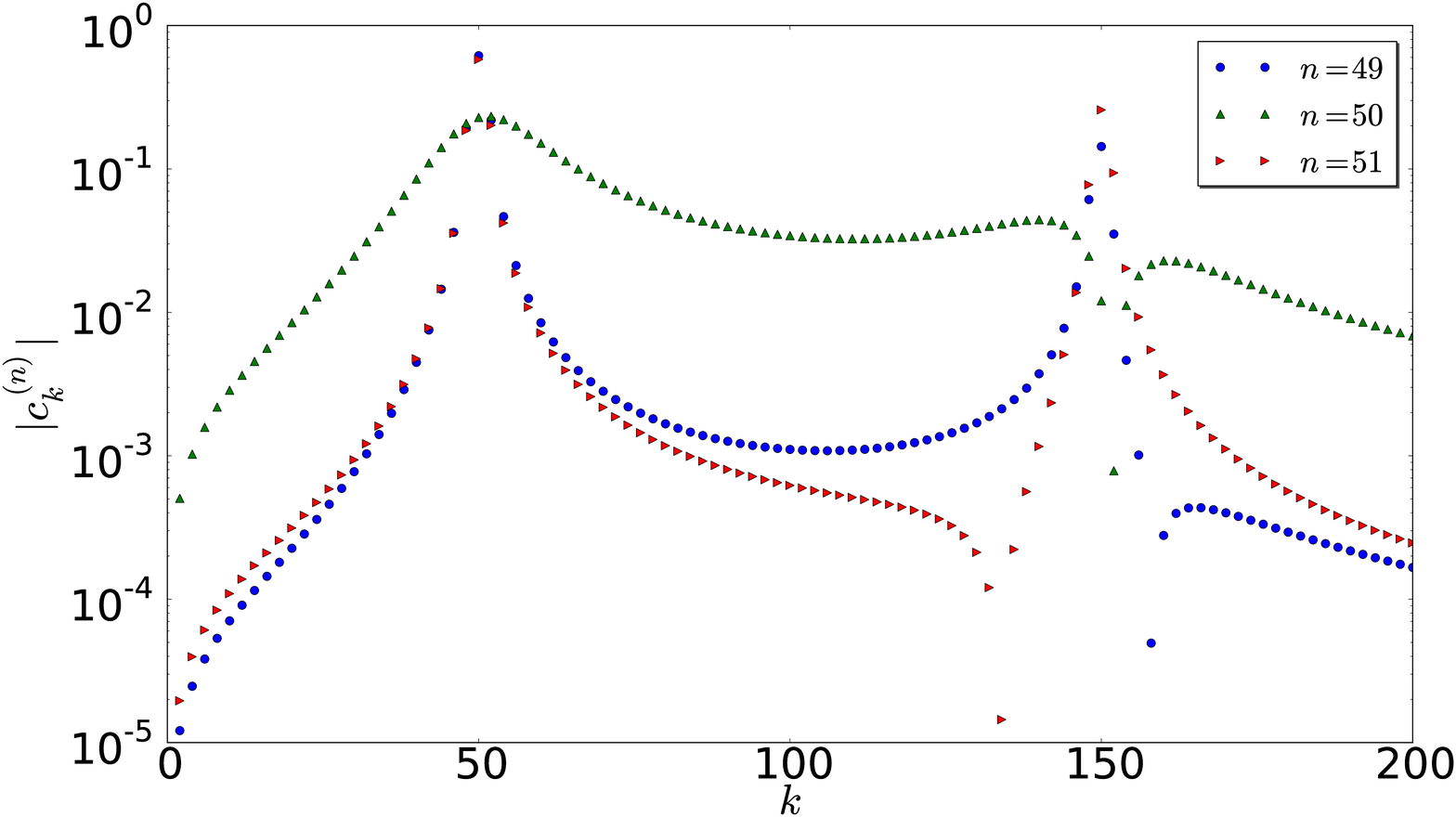}
\caption{Approximate even Fourier coefficients for the modes $49$,$50$ and $51$ of the string with density 
$\rho(x)=\left(2+ \sin (100 \pi  x)\right)^2$ obtained with LSF with a grid of $2500$ points. }
\label{Fig_7even}
\end{center}
\end{figure}

In Fig.~\ref{Fig_7odd} and \ref{Fig_7even} we display the approximate Fourier coefficients for the solutions of Fig.\ref{Fig_6} for 
$n$ odd and even respectively, obtained with LSF on a grid of $2500$ points. The coefficients of the solutions corresponding
to $n=49$ and $n=51$ behave very differently for odd and even modes while the coefficients of the localized solution, 
corresponding to $n=50$, behave similarly in the two cases and decay slowly.

We may now discuss the bounds for the energy of the fundamental mode obtained using the different approaches 
outlined in the previous sections; in the case of DPT one uses the fundamental solution of a uniform string 
as variational ansatz and obtains a bound which falls quite above the exact energy:
\beq
E_1 \leq 6335.15 \ .
\eeq

The situation is only partially better using WKBPT: the "potential" $V(x)$ corresponding to this density is
\beq
V(x) = -\frac{1250 \pi ^2 (8 \sin (100 \pi  x)+\cos (200 \pi  x)+5)}{(\sin (100 \pi  x)+2)^4}
\eeq
and using the formula derived earlier the bound reads
\beq
E_1 \leq 1189.60 \ .
\eeq

This large bound originates almost entirely from the first order contribution
\beq
E_1^{(1)} = \frac{2}{\sigma(L)} \int_{-L}^{L} \sqrt{\rho(x)} V(x) \sin^2 \frac{\pi \sigma(x)}{\sigma(L)} dx 
\approx  \frac{1}{\sigma(L)} \int_{-L}^{L} \sqrt{\rho(x)} V(x) dx \ .
\eeq

This result clearly suggests that the WKBPT describes poorly the lowest part of the spectrum of this system, 
although the basis of the eigenfunctions of $\hat{Q}$ reproduces the asymptotic behavior of the high part of 
the spectrum of $\hat{O}$. Both bounds calculated so far are clearly useless!

In the iWKBPT approach one may use an effective density $\tilde{\rho}(x)$, different from the physical density
and specific to a given state, to improve the description of the lowest energy solutions. 
As we have seen in the case of WKBPT discussed above, the large values obtained in the bound originate from the 
rapid oscillations of the density: one should therefore pick an effective density $\tilde{\rho}(x)$ for which 
these contributions are suppressed.

An useful ansatz is 
\beq
\tilde{\rho}(x) \approx \rho(x)^2 \ ,
\eeq
since it cancels large contributions to the expectation value of $\hat{O}$ stemming from the oscillations
of the density.

The bound for the fundamental energy obtained in this case is much closer to the numerical result obtained 
with collocation
\beq
E_1 \leq 3.07325 \ ,
\eeq
although the quality of this bound is not competing with the precision of the numerical result.

We will now show that it is possible to obtain extremely precise variational bounds, choosing a suitable variational ansatz,
given by the trial function (normalized to one)
as
\beq
\psi_1^{(trial)}(x) = \frac{\sqrt{2}}{3} \sqrt{\rho(x)/L} \sin\left[\pi \frac{x+L}{2L}\right] \ .
\label{trial}
\eeq

A simple calculation shows that the bound is this case is
\beq
E_1 \leq \frac{2}{9} \pi^2 \approx 2.19325 \ ,
\eeq
where the first four digits of the rhs are correct. 

Can we still improve this bound? Yes, we can: we can use Theorem 1 of ref.~\cite{Amore10b} 
to obtain further analytical improvements of this bound. 
Using the notation of that paper we set $\xi_0(x) \equiv \psi_1^{(trial)}(x)$ and obtain refined 
approximations for the fundamental solution applying the iterative equation eq.~(\ref{iter}).

The bounds obtained in this way are given in Table \ref{table:6}: after $7$ iterations it appears that the
first $19$ digits have converged. In principle the iterations can be performed to arbitrary orders, as long as
one is able to perform the integrals analytically~\footnote{The form of the density in this example was chosen 
to allow to perform several iterations.}. A comparison with the LSF result shows that the first 10 
digits of the value obtained with collocation are correct. 

\begin{table}[tbp]
\caption{Variational bounds for the energy of the fundamental mode of the string of unit length of density
$\rho(x)=\left(2+ \sin (100 \pi  x)\right)^2$ obtained using Theorem 1 of ref.~\cite{Amore10b} with the initial
trial function in eq.~(\ref{trial}).}
\bigskip
\label{table:6}
\begin{center}
\begin{tabular}{|c|c|}
\hline
$n$ &    $E_1$ \\
\hline
0   &     2.1932454224643019153\\
1   &     2.1931584087453807530\\
2   &     2.1931584087440065685\\
3   &     2.1931584087439513989\\
4   &     2.1931584087439479612\\
5   &     2.1931584087439477463\\
6   &     2.1931584087439477329\\
7   &     2.1931584087439477320\\
\hline
\end{tabular}
\end{center}
\bigskip\bigskip
\end{table}

\begin{figure}
\begin{center}
\bigskip\bigskip\bigskip
\includegraphics[width=11cm]{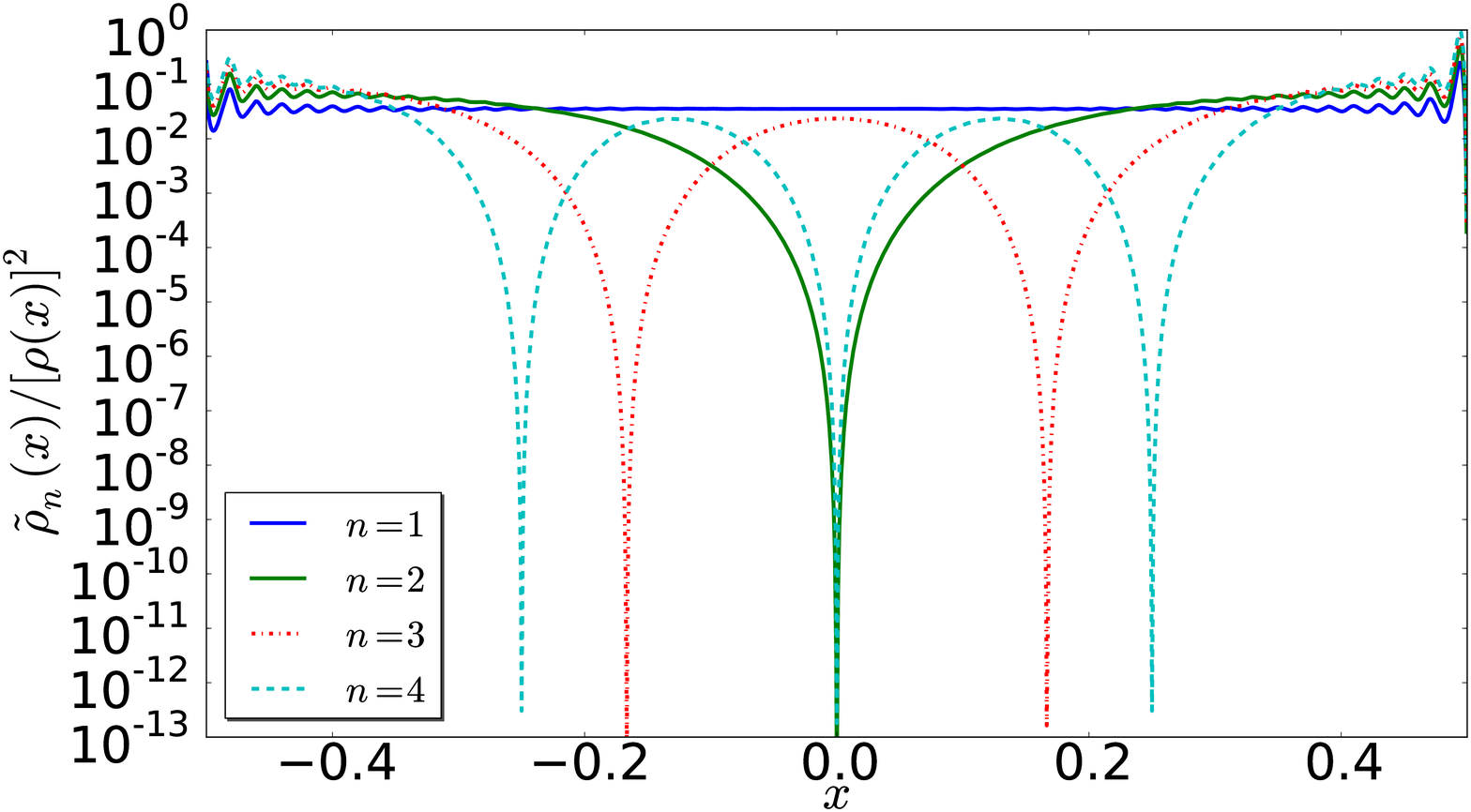}
\caption{Effective densities for the first four solutions of the string with density 
$\rho(x)=\left(2+ \sin (100 \pi  x)\right)^2$ obtained with LSF with a grid of $2500$ points. }
\label{Fig_rho}
\end{center}
\end{figure}

Having obtained accurate numerical solutions for the low lying states of this string, we may look for the 
effective densities corresponding to these states. In this case one needs to solve numerically the 
equation
\beq
\frac{\tilde{\sigma}(x)}{\tilde{\sigma}(L)} &=& \frac{1}{2\pi} \sin \frac{2\pi \tilde{\sigma}(x)}{\tilde{\sigma}(L)} + 
\int_{-L}^x \left[\psi_1^{(trial)}(y)\right]^2 dy \ .
\label{selfcons}
\eeq
The effective density is then obtained through the relation $\tilde{\sigma}'(x) =\sqrt{\tilde{\rho}(x)}$.
In Fig.\ref{Fig_rho} we show the effective densities for the first four solutions, scaled by the factor
$\rho(x)^2$, which we have already seen to be a good approximation: as a matter of fact we may notice that
$\tilde{\rho}_1(x)/\rho^2(x)$ is almost flat, with oscillations only at the ends of the string. 

In the case of the remaining states we notice a very peculiar behavior of the effective density, which 
develops $n-1$ zeroes in the domain, which also coincide with the zeroes of the solution (the zeroes
of the eigenfunctions of $\tilde{\hat{Q}}$  correspond either to the zeroes of the trigonometric function or
to the zeroes of the effective density, as in the present case).

\begin{figure}
\begin{center}
\bigskip\bigskip\bigskip
\includegraphics[width=11cm]{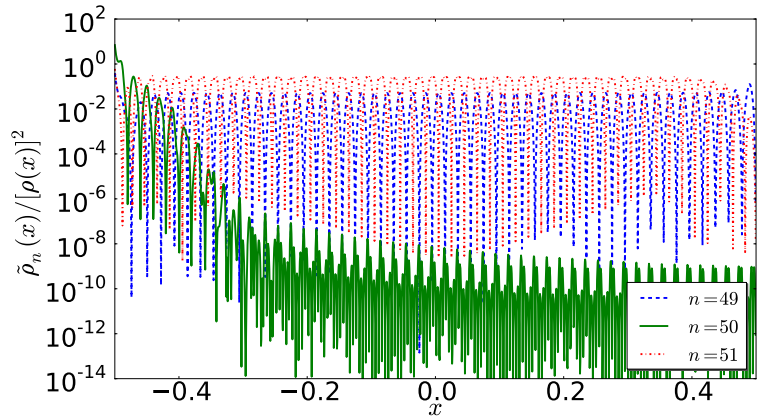}
\caption{Effective densities for the solutions $49$, $50$ and $51$ of the string with density 
$\rho(x)=\left(2+ \sin (100 \pi  x)\right)^2$ obtained with LSF with a grid of $2500$ points. }
\label{Fig_rho1}
\end{center}
\end{figure}

In Fig.~\ref{Fig_rho1} we display the same ratio for the modes corresponding to $n=49$, $50$ and $51$. 
The effective density of the mode $50$ is clearly localized at the left end of the string.

\section{Low  and high energy asymptotics for strings with rapidly oscillating density}

In this section we use the theorems of Section \ref{sec:iterative} to obtain the low energy asymptotics of a string with 
density~\footnote{The low energy asymptotics for this problem has been obtained by Castro and Zuazua in \cite{CZ00b}. }
\beq
\rho(x) &=& 2+\sin \left(\frac{2 \pi  \left(x+\frac{1}{2}\right)}{\epsilon }\right)
\label{densosc}
\eeq
with $\epsilon \rightarrow 0^+$ and $x^2 \leq 1/4$. 

As we have seen through Theorem \ref{theo1} it is possible to obtain a sequence of functions
which converge to the fundamental solution of $\hat{O}$ provided that the initial trial function
is not orthogonal to it. Therefore we may consider the function
\beq
\xi_0(x) = \sqrt{\rho(x)} \sin \frac{\pi (x+1/2)}{2L}  \ .
\eeq

Actually this $\xi_0(x)$ not only is not orthogonal
to the exact fundamental mode of $\hat{O}$, but it also tends to become an exact eigenfunction of $\hat{O}$ 
in the limit $\epsilon \rightarrow 0^+$.

We can see this by calculating 
\beq
\frac{\hat{O} \xi_0(x)}{\xi_0(x)}  = \frac{\pi ^2}{2+\sin \left(\frac{2 \pi  \left(x+\frac{1}{2}\right)}{\epsilon }\right)}
\approx \frac{\pi^2}{2}  \ ,
\eeq
having substituted in the last expression the highly oscillatory term 
$\sin \left(\frac{2 \pi  \left(x+\frac{1}{2}\right)}{\epsilon }\right)$ with its average value on the interval.

We thus use Theorem \ref{theo1} to build increasingly refined approximations of the fundamental solution of $\hat{O}$ corresponding 
to the density (\ref{densosc}) in the limit $\epsilon \rightarrow 0^+$. For instance, after one iteration we obtain the
function:
\beq
\xi_1(x) &=& \frac{\sqrt{\sin \left(\frac{2 \pi  \left(x+\frac{1}{2}\right)}{\epsilon }\right)+2}}{2 \pi^2
   \left(\epsilon^2-4\right)^2} \ \left[4 \epsilon^3 (2 x+1) \cos \left(\frac{2 \pi }{\epsilon }\right)  \right. \nonumber \\
&+& \left.\epsilon ^2 \left(4 \epsilon  (2x-1)+(\epsilon -2)^2 \sin \left(\frac{\pi  ((\epsilon +2) x+1)}{\epsilon }\right) \right.\right. \nonumber \\
&+& \left.\left.(\epsilon +2)^2\sin \left(\frac{\pi  (-\epsilon  x+2 x+1)}{\epsilon }\right)\right) +4 \left(\epsilon^2-4\right)^2 \cos (\pi  x)\right] \ .
\eeq

We have also calculated explicitly the functions corresponding to two and three iterations, although their expressions
are much lengthier and therefore we will not report them here.

We will instead calculate the expectation value of $\hat{O}$ in these functions, thus obtaining an asymptotic
expression for the fundamental energy of this string in the limit $\epsilon \rightarrow 0^+$. 

We report here these expressions:
\beq
\langle \hat{O} \rangle_{\xi_0} &\equiv& \frac{\int_{-L}^{+L} \xi_0(x) \hat{O} \xi_0(x) dx}{\int_{-L}^{+L} \xi_0(x)^2 dx} \approx
\frac{\pi ^2}{2} + \frac{1}{4} \pi  \sin ^2\left(\frac{\pi }{\epsilon }\right) \epsilon^3 + 
\frac{1}{4} \pi  \sin ^2\left(\frac{\pi }{\epsilon }\right) \epsilon^5 + \dots \\
\langle \hat{O} \rangle_{\xi_1} &\equiv& \frac{\int_{-L}^{+L} \xi_1(x) \hat{O} \xi_1(x) dx}{\int_{-L}^{+L} \xi_1(x)^2 dx} \approx 
\frac{\pi ^2}{2} -\frac{\pi ^2}{64} \epsilon^2 + \frac{1}{4} \pi  \sin ^2\left(\frac{\pi }{\epsilon }\right) \epsilon^3
-\frac{15 \pi ^2}{1024} \epsilon^4 \nonumber \\
&+& \frac{\pi  \left(5 \sin \left(\frac{4 \pi }{\epsilon }\right)-116 \cos \left(\frac{2 \pi }{\epsilon
   }\right)+116\right)}{1024} \epsilon^5 + \dots \\
\langle \hat{O} \rangle_{\xi_2} &\equiv& \frac{\int_{-L}^{+L} \xi_2(x) \hat{O} \xi_2(x) dx}{\int_{-L}^{+L} \xi_2(x)^2 dx} \approx 
\frac{\pi^2}{2} -\frac{\pi ^2}{64} \epsilon^2 + \frac{1}{4} \pi  \sin ^2\left(\frac{\pi }{\epsilon }\right) \epsilon^3
-\frac{15 \pi ^2}{1024} \epsilon^4 \nonumber \\
&+& \frac{\pi  \left(5 \sin \left(\frac{4 \pi }{\epsilon }\right)-116 \cos \left(\frac{2 \pi }{\epsilon
   }\right)+116\right)}{1024} \epsilon^5 + \dots \\ 
\eeq

As we can see the expectation value of $\hat{O}$ has converged to order $\epsilon^5$ after just one iteration; thus we can 
conclude that, for $\epsilon \rightarrow 0^+$, the energy of the fundamental mode goes like
\beq
E_1 &\approx& \frac{\pi^2}{2} -\frac{\pi ^2}{64} \epsilon^2 + \frac{1}{4} \pi  \sin ^2\left(\frac{\pi }{\epsilon }\right) \epsilon^3
-\frac{15 \pi ^2}{1024} \epsilon^4 \nonumber \\
&+& \frac{\pi  \left(5 \sin \left(\frac{4 \pi }{\epsilon }\right)-116 \cos \left(\frac{2 \pi }{\epsilon
   }\right)+116\right)}{1024} \epsilon^5 + O\left[\epsilon^6\right] \ .
\eeq

This result may be compared with eq.(3.33) of Ref.\cite{CZ00b}, which reports $\sqrt{E_1}$ to order $O\left[\epsilon^4\right]$ for 
this example.

Let us now derive the asymptotics  for the low energy states of this string; in this case we use Theorem \ref{theo3} 
and consider the trial function
\beq
\eta_0(x) = \sqrt{2+\sin \left(\frac{2 \pi  \left(x+\frac{1}{2}\right)}{\epsilon }\right)}  \ \sin \left(\frac{\pi 
   \left(x+\frac{1}{2}\right)}{{\bar{L}}+\frac{1}{2}}\right) \ ,
\eeq
with $\eta_0(-L) = \eta_0(\bar{L})=0$. Notice that $\eta_0(L) = 0$ only for $\bar{L} = - \frac{1}{2} + \frac{1}{n}$, with $n = 1,2, \dots$
and
\beq
\frac{\hat{O} \eta_0(x)}{\eta_0(x)}  = \frac{n^2\pi^2}{2 + \sin \left(\frac{2 \pi  x+\pi }{\epsilon }\right)} 
\approx \frac{n^2 \pi^2}{2}  \ ,
\eeq
meaning that for $\epsilon \rightarrow 0^+$, the $\eta_0(x)$ for $\bar{L} = - \frac{1}{2} + \frac{1}{n}$ converges to the
$n^{th}$ solution for this string (again we are substituting the oscillating factor with its average value on the 
interval in this limit).

We may now apply Theorem \ref{theo3}: after one iteration we obtain the function
\beq
\eta_1(x) &=& \frac{(2 {\bar{L}}+1)^2 \sqrt{\sin \left(\frac{2 \pi  \left(x+\frac{1}{2}\right)}{\epsilon}\right)+2}}{8 
\pi ^2 (\epsilon -2 {\bar{L}}-1)^2 (\epsilon +2 {\bar{L}}+1)^2} \left[ 4 \epsilon ^3 (2 x+1) \cos \left(\frac{2 \pi  {\bar{L}}+\pi }{\epsilon }\right)
\right. \nonumber \\
&+& \left.\epsilon ^2 (\epsilon
   +2 {\bar{L}}+1)^2 \cos \left(\frac{\pi  (2 x+1) (\epsilon -2 {\bar{L}}-1)}{2 \epsilon {\bar{L}}+\epsilon }\right) \right. \nonumber \\
&+& \left. \epsilon ^2 (\epsilon -2 {\bar{L}}-1)^2 \cos \left(\frac{\pi ({\bar{L}} (-2 \epsilon +4 x+2)+2 (\epsilon +1) x+1)}{2 \epsilon  {\bar{L}}+\epsilon }\right) \right. \nonumber \\
&+& \left. 4 \left(2 \epsilon ^3 (x-{\bar{L}})+\left(\epsilon ^2-(2 {\bar{L}}+1)^2\right)^2 \sin \left(\frac{2
   \pi  x+\pi }{2 {\bar{L}}+1}\right)\right) \right] \ .
\label{eta1_theo3}
\eeq

The zeroes of the equation
\beq
\eta_1(L) = 0
\eeq
will now provide improved values for $\bar{L}$ and for the energies.

\begin{figure}
\begin{center}
\bigskip\bigskip\bigskip
\includegraphics[width=11cm]{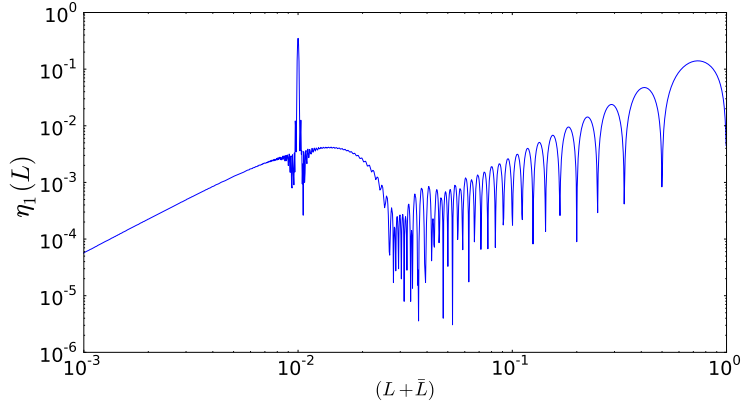}
\caption{The function of eq.(\ref{eta1_theo3}) evaluated at $x=L$, for $\epsilon = 1/50$ as a function
of $(L+ \bar{L})$.}
\label{Fig_theo3}
\end{center}
\end{figure}

In Fig.\ref{Fig_theo3} we display $\eta_1(L)$ corresponding to $\epsilon = 1/50$, as a function of $(L+ \bar{L})$: the 
spikes of the curve correspond to zeroes of $\eta_1(L)$, located at specific values of $\bar{L}$. Corresponding to each value 
Theorem 3 provides an approximate eigenvalue and eigensolution to the string problem.

In Table \ref{table:7} we report the values of $\bar{L}$ corresponding to the first 20 zeroes of $\eta_1(L)$ and the
corresponding energies obtained by calculating the expectation value of $\hat{O}$ in the corresponding state; the
last column contains the precise numerical results obtained with LSF collocation with a grid of $2500$ points. Thus
we see that Theorem 3 provides rather precise results for a large number of states just after one iteration. 

\begin{table}[tbp]
\caption{Approximate energies of a string with density (\ref{densosc}) with $\epsilon=1/50$ obtained with Theorem \ref{theo3} 
(third column) and corresponding values obtained with LSF on a grid of $2500$ points (fourth column). The second column
contains the values of $\bar{L}$ for which $\eta_1(L)=0$ (see the figure).}
\bigskip
\label{table:7}
\begin{center}
\begin{tabular}{|c|c|c|c|}
\hline
$n$ & $\bar{L}$ &    $E_n$ & $E_n^{{\rm LSF_{2500}}}$ \\
\hline
1  &  0.50000000000000 & 4.93474049237786 & 4.93474049164385 \\
2  &  0.00000254847810 & 19.7382203568987 & 19.7382203589595 \\
3  & -0.16666618871377 & 44.4082063869071 & 44.4082063970937 \\
4  & -0.25000000000000 & 78.9409479076410 & 78.9409485235574 \\
5  & -0.30001279890522 & 123.331046653456 & 123.331137037346 \\
6  & -0.33332758093170 & 177.571827239743 & 177.571841516164 \\
7  & -0.35716402051048 & 241.652567358357 & 241.654430048229 \\
8  & -0.37497950785717 & 315.566145317546 & 315.568467036379 \\
9  & -0.38888465796479 & 399.301373439805 & 399.301587271749 \\
10 & -0.39993517619554 & 492.724112367784 & 492.839343430514 \\
11 & -0.40908535234753 & 596.163807132223 & 596.165023435926 \\
12 & -0.41659695736991 & 708.868637232727 & 709.259433238896 \\
13 & -0.42315478442053 & 831.113631633346 & 832.100639666984 \\
14 & -0.42856494498590 & 964.660849364111 & 964.663666640351 \\
15 & -0.43336143492296 & 1106.57429450867 & 1106.92013667688 \\
16 & -0.43735701119023 & 1249.09849824817 & 1258.83784766743 \\
17 & -0.44135393715531 & 1396.14689996551 & 1420.38027258297 \\
18 & -0.44431129182849 & 1574.93187010870 & 1591.50596694241 \\
19 & -0.44739462363259 & 1770.83890618558 & 1772.16786511172 \\
20 & -0.45000000000000 & 1962.04796373469 & 1962.31244196980 \\
\hline
\end{tabular}
\end{center}
\bigskip\bigskip
\end{table}

We look for approximate solutions of the form
\beq
\bar{L} = \left(- \frac{1}{2} + \frac{1}{n}\right) + \delta \bar{L}
\eeq
where $\delta \bar{L}$ is a correction. To leading order we have
\beq
\delta \bar{L} \approx \frac{\epsilon ^3 (-1)^{-n} n \left(n \cos \left(\frac{2 \pi }{\epsilon  n}\right)+(-1)^n \cos
   \left(\frac{2 \pi }{\epsilon }\right)+n-1\right)}{8 \pi } \ .
\eeq
Notice that $\delta \bar{L}=0$ for $n=1$, as it should be.

Working to order $\epsilon^3$ we have calculated the expectation value of $\hat{O}$ in $\eta_1(x)$ to order $\epsilon^3$:
\beq
\langle \hat{O} \rangle_{\eta_1} &\equiv& \frac{\int_{-L}^{\bar{L}} \eta_1(x) \hat{O} \eta_1(x) dx}{\int_{-L}^{\bar{L}} \eta_1(x)^2 dx} \approx 
\frac{\pi ^2 n^2}{2} -\frac{1}{64} \pi ^2 \epsilon ^2 n^4 + \frac{\epsilon^3}{4} \pi  n^4 \sin ^2\left(\frac{\pi }{\epsilon }\right)
+ O\left[\epsilon^4\right] \nonumber \ ,
\eeq
which provides the asymptotic behavior of the energy of this string for $\epsilon \rightarrow 0^+$. Notice that
\beq
\sqrt{E_n} \approx \frac{\pi  n}{\sqrt{2}}  -\frac{\pi \epsilon ^2 n^3}{64 \sqrt{2}}
 + \frac{\epsilon ^3 n^3 \sin ^2\left(\frac{\pi }{\epsilon }\right)}{4 \sqrt{2}}
+ O\left[\epsilon^4\right] \nonumber \ ,
\eeq
coincides with the result obtained by Castro and Zuazua in ref.~\cite{CZ00b} using the WKB method. The advantage of the present
approach is that the calculation of the higher order asymptotics only requires to apply further iterations and 
expand the expectation value of $\hat{O}$ to higher order (of course one needs to be able to perform the iterations explicitly!). 
Moreover, the theorem \ref{theo3} allows to obtain at the same time approximations for the energies and for the solutions.

We will now discuss the high energy asymptotics for this string, using the asymptotic expansion obtained with WKBPT.
A simple calculation shows that in the present case:
\beq
V(x) = -\frac{\pi ^2 \left(16 \sin \left(\frac{2 \pi  x+\pi }{\epsilon }\right)+\cos \left(\frac{2 \pi  (2 x+1)}{\epsilon }\right)+9\right)}{8 \epsilon ^2 \left(\sin \left(\frac{2 \pi  x+\pi }{\epsilon }\right)+2\right)^3} \ .
\eeq

We also find that
\beq
\sigma(x) = \int_{-L}^x \sqrt{\rho(y)} dy = \frac{\sqrt{3} \epsilon  \left(E\left(\frac{\pi }{4}|\frac{2}{3}\right)-E\left(\frac{\pi  (\epsilon -4 x-2)}{4 \epsilon }|\frac{2}{3}\right)\right)}{\pi } \ ,
\eeq
where $E(x | m) \equiv \int_0^x  \sqrt{1-m \sin^2 t} \ dt$ is the incomplete elliptic integral of second kind.

Notice that
\beq
\lim_{\epsilon\rightarrow 0^+} \sigma(L) =  \frac{2 \sqrt{3}}{\pi }  \ E\left(\frac{2}{3}\right) \approx 1.39066 \ .
\eeq

The dominant contribution to the high energy asymptotics of this string is readily obtained using these expressions:
\beq
E_n \approx \frac{\pi n}{\sigma(L)} + \dots \approx 2.26 n + \dots \ ,
\eeq
for $n \rightarrow \infty$.

We come now to the subleading term, which requires the calculation of
\beq
\langle V \rangle &=& \frac{1}{\sigma(L)} \int_0^{\sigma(L)} V(\sigma(x)) d\sigma \nonumber \\
&=& -\frac{23 \pi }{72 \sqrt{2} \epsilon  \sigma(L) }+\frac{\pi  \sin \left(\frac{4 \pi }{\epsilon }\right)}{18 \epsilon  \sigma(L)  \left(\sin \left(\frac{2 \pi }{\epsilon }\right)+2\right)^{3/2}}
    +\frac{\pi F\left(\frac{(\epsilon -4) \pi }{4 \epsilon }|\frac{2}{3}\right)}{6 \sqrt{3} \epsilon  \sigma(L) } \nonumber \\
&-& \frac{\pi  E\left(\frac{(\epsilon -4) \pi }{4 \epsilon }|\frac{2}{3}\right)}{3 \sqrt{3} \epsilon 
   \sigma(L) }+\frac{\pi  E\left(\frac{\pi }{4}|\frac{2}{3}\right)}{3 \sqrt{3} \epsilon  \sigma(L) }+\frac{23 \pi  \cos \left(\frac{2 \pi }{\epsilon }\right)}{36 \epsilon  \sigma(L)  \left(\sin \left(\frac{2 \pi
   }{\epsilon }\right)+2\right)^{3/2}} \nonumber \\
&-& \frac{\pi  F\left(\csc ^{-1}\left(\sqrt{3}\right)|\frac{3}{2}\right)}{6 \sqrt{2} \epsilon  \sigma(L) } \ ,
\eeq
where $F(x | m) \equiv \int_0^x  \frac{1}{\sqrt{1-m \sin^2 t}} \ dt$ is the incomplete elliptic integral of first kind.

Notice that $\epsilon^2 \langle V \rangle $ tends to a constant as $\epsilon \rightarrow 0^+$: 
$\lim_{\epsilon\rightarrow 0^+} \langle V \rangle = \kappa$.
Therefore, for a given $\epsilon$, arbitrarily small and positive, the behavior of the energy will be dominated by the leading term
only for states with $n \gg \kappa \sigma(L)/\pi\epsilon^2$.
\beq
\lim_{\epsilon \rightarrow 0^+} \epsilon^2 \langle V\rangle = \frac{\pi^2}{18}  \left(2-\frac{K\left(\frac{2}{3}\right)}{E\left(\frac{2}{3}\right)}\right)
\approx 0.214515  \ ,
\eeq
where $K(m)$ and $E(m)$ are the complete elliptic integrals of first and second kind respectively.

\section{Conclusions}
\label{sec:conclusions}

In this paper we have devised an alternative perturbative approach to the problem of finding the normal modes of
a string with arbitrary density: using this approach we have obtained an asymptotic high energy expansion, and we have
derived explicit analytical expression for the first three coefficients of this expansion. The first two coefficients 
were already known and they can be derived using the WKB method; the third coefficient was not known (to the best of our
knowledge) and it is expressed in terms of a series whose summands are themself series (divergent ones), which can be 
evaluated using a Borel transform. Higher order coefficients could also be calculated in a similar fashion, although 
we have not done it here. We have discussed specific cases, where exact solutions (or very precise numerical solutions)
were available, and we have reproduced the theoretical values obtained with our formula with high  accuracy.

We have also applied the iterative theorems of ref.~\cite{Amore10b} to the problem of a string with highly oscillatory
density; in this way we have obtained an explicit analytical formula which describes the low energy asymptotics of a
string with density of arbitrarily small wavelength and we have thus reproduced the results of Castro and Zuazua 
\cite{CZ00b}, who had used the WKB method.

We believe that several issues could be investigated using the techniques described in this paper and in ref.\cite{Amore10b}; 
in particular, we plan to calculate in a future work few more higher order asymptotic coefficients for the energies of
strings with arbitrary density using the WKBPT method of the present paper and to perform a similar calculation
of the asymptotic form of the solutions, which has not been carried out here.

Finally, we would like to bring the attention of the reader to a "pedagogical" aspect of the perturbative scheme devised
in the present article: while both perturbation theory and the WKB method are a standard part of most quantum 
mechanics/mathematical physics books, normally they are discussed separately as unrelated topics. The WKB-perturbation
method that we have described here shows that it is possible to merge the two approaches in a single and more powerful 
method, at least for the problem of calculating the normal modes of inhomogeneous strings of arbitrary density.
Its extension to higher dimensional problems is certainly not trivial, although it deserves to be investigated in the future.

\appendix

\section{Asymptotic expressions for the the matrix elements}
\label{sec:appendix:1}

In this appendix we work out the asymptotic expressions for the matrix elements $\langle n | V | k\rangle$, which
appear in the WKBPT perturbative series.

We start with the diagonal terms, which appear in the first order correction $E_n^{(1)}$:
\beq
\langle n | V | n \rangle &=& \frac{2}{\sigma(L)} \int_0^{\sigma(L)} \sin^2 \left(\frac{n \pi \sigma}{\sigma(L)}\right) \ 
V(x(\sigma)) \ d\sigma \nonumber \\
&=& \langle V \rangle  - \frac{1}{\sigma(L)} \ \int_0^{\sigma(L)}  \cos \left(\frac{2 n \pi \sigma}{\sigma(L)}\right)  
V(x(\sigma)) \ d\sigma 
\label{sec:app:1}
\eeq
where
\beq
\langle V \rangle \equiv \frac{1}{\sigma(L)} \ \int_0^{\sigma(L)} V(x(\sigma)) \ d\sigma 
\label{sec:app:2}
\eeq
is the average of $V$ on the interval $(0,\sigma(L))$.

The second term may be evaluated using the integration by part: 
\beq
{\rm (II)} &=& - \frac{1}{\sigma(L)} \ \int_0^{\sigma(L)}  \cos \left(\frac{2 n \pi \sigma}{\sigma(L)}\right)  V(x(\sigma)) \ 
d\sigma \nonumber \\ 
&=& - \frac{1}{2 n \pi} \ \int_0^{\sigma(L)}  \sin \left(\frac{2 n \pi \sigma}{\sigma(L)}\right)  \frac{dV(x(\sigma))}{d\sigma} \ 
d\sigma \nonumber \\
&=& \frac{\sigma(L)}{4 \pi^2 n^2} \left[  \left.\frac{dV(x(\sigma))}{d\sigma}\right|_{\sigma(L)} -\left.\frac{dV(x(\sigma))}{d\sigma}
\right|_{0} \right] \nonumber \\
&-& \frac{\sigma(L)}{4 \pi^2 n^2} \int_0^{\sigma(L)} \cos \left(\frac{2 n \pi \sigma}{\sigma(L)}\right)  
\frac{d^2V(x(\sigma))}{d\sigma^2} \ d\sigma \ .
\label{sec:app:3}
\eeq

Since the last term in this expression has a form which is similar to that of the expression that we started with, the 
integration may be obtained with no effort:
\beq
{\rm (II)} &=&  \frac{1}{\sigma(L)} \ \int_0^{\sigma(L)}  \cos \left(\frac{2 n \pi \sigma}{\sigma(L)}\right)  V(x(\sigma)) \ d\sigma 
\nonumber \\ 
&=& \sum_{k=0}^\infty \frac{\sigma(L)^{2k+1}}{(2 \pi n)^{2k+2}} (-1)^k \left[ V^{(2k+1)}(\sigma(L))  - V^{(2k+1)}(0) \right] \ .
\label{sec:app:4}
\eeq

Therefore
\beq
E_n^{(1)} = \langle V \rangle -\sum_{k=0}^\infty \frac{\sigma(L)^{2k+1}}{(2 \pi n)^{2k+2}} (-1)^k \left[ V^{(2k+1)}(\sigma(L))  - V^{(2k+1)}(0) 
\right] \ .
\eeq

The same strategy may be followed for the non-diagonal matrix elements; after performing repeated integrations by part one
is left with the expression:
\beq
\langle n | V | k \rangle &=& \frac{2}{\sigma(L)} \int_0^{\sigma(L)} \sin\left(\frac{n \pi \sigma}{\sigma(L)}\right) 
\sin\left(\frac{k \pi \sigma}{\sigma(L)}\right) \ V(x(\sigma)) \ d\sigma \nonumber \\
&=& \frac{1}{\sigma(L)} \int_0^{\sigma(L)} \left[ \cos\left(\frac{(k-n)\pi \sigma}{\sigma(L)} \right)
- \cos\left(\frac{(k+n)\pi \sigma}{\sigma(L)} \right) \right] V(x(\sigma)) d\sigma \nonumber \\
&=& \sum_{j=0}^\infty (-1)^j \frac{\sigma(L)^{2j+1}}{\pi^{2j+2}} \left(\frac{1}{(k-n)^{2j+2}}-\frac{1}{(k+n)^{2j+2}} \right)
\nonumber \\
&\cdot& \left[ (-1)^{k+n} V^{(2j+1)}(\sigma(L))-V^{(2j+1)}(0)\right]
\eeq

Notice that:
\begin{itemize}
\item $\lim_{n\rightarrow \infty} \langle n | V | n \rangle = \langle V \rangle$;
\item $\lim_{n\rightarrow \infty} \langle n | V | k \rangle = 0$ for $k\neq n$;
\end{itemize}

As a result, $E_n^{(1)}$ dominates all the remaining terms in the perturbative series as $n \rightarrow \infty$, as it 
should be. In particular, for $n \gg 1$ the non--diagonal matrix element behaves as
\beq
\langle n | V | k \rangle \approx \frac{\sigma(L)}{\pi^2} \ \frac{4 kn}{(k^2-n^2)^2}  \left[ (-1)^{k+n} V^{(1)}(\sigma(L))-V^{(1)}(0)
\right] \ .
\eeq

\section{WKB approximation}
\label{sec:appendix:2}

In this appendix we describe the application of the WKB approximation to the problem of a string of variable density and 
derive the first few corrections to the energies of the modes of the string. Our discussion follows closely ref.~\cite{BO78},
where however only the leading contribution was derived explicitly.

Our starting point is eq.~(\ref{sec:dpt:1}): we assume
\beq
\Psi_n(x) \approx e^{\frac{1}{\delta} \sum_{n=0}^\infty \delta^n S_n(x)} \ ,
\label{sec:app2:1}
\eeq
with $\delta \rightarrow 0$. Assuming $E_n$ to be of order $1/\delta^2$, eq.~(\ref{sec:dpt:1}) may be written
as
\beq
- \delta^2 \frac{d^2\Psi_n(x)}{dx^2} = E_n \rho(x) \Psi_n(x) \ .
\label{sec:app2:2}
\eeq

Substituting eq.~(\ref{sec:app2:1}) inside eq.~(\ref{sec:app2:2}) we obtain the equations for the different orders in $\delta$.
To leading order, for example, the equation obtained is
\beq
S_0'(x)^2 + E_n \rho(x) = 0 
\eeq
whose solution is 
\beq
S_0(x) = \pm i \sqrt{E_n} \int_{-L}^x \sqrt{\rho(y)}  dy  \ .
\eeq

To first order we have
\beq
-2 S_0'(x) S_1'(x) - S_0''(x) =0  \ ,
\eeq
whose solution is 
\beq
S_1(x) =  - \frac{1}{4} \log \rho(x) + c
\eeq
where $c$ is a constant of integration, whose value should be fixed by the normalization of the solution.

To second order we have
\beq
- S_1'(x)^2 -2 S_0'(x) S_2'(x) - S_1''(x) = 0
\eeq
whose solution is
\beq
S_2(x) &=& - \int_{-L}^x \frac{S_1'(y)^2+S_1''(y)}{2 S_0'(y)} dy \nonumber \\
&=& \mp \frac{i}{32 \sqrt{E_n}} \int_{-L}^x \frac{ 5 \rho'(y)^2-4 \rho(y) \rho''(y)}{\rho(y)^{5/2}} dy \ .
\eeq

Substituting these results in eq.(\ref{sec:app2:1}) we have
\beq
\Psi_n(x) \propto \frac{1}{\rho(x)^{1/4}} \ e^{\pm i  \sqrt{E_n} \int_{-L}^x \sqrt{\rho(y)}  dy
\mp \frac{i}{32 \sqrt{E_n}} \int_{-L}^x \frac{ 5 \rho'(y)^2-4 \rho(y) \rho''(y)}{\rho(y)^{5/2}} dy + \dots}
\eeq
or equivalently
\beq
\Psi_n(x) &\propto& \frac{1}{\rho(x)^{1/4}} \ \sin\left[\sqrt{E_n} \int_{-L}^x \sqrt{\rho(y)}  dy \right. \nonumber \\
&-& \left. \frac{1}{32 \sqrt{E_n}} \int_{-L}^x \frac{ 5 \rho'(y)^2-4 \rho(y) \rho''(y)}{\rho(y)^{5/2}} dy + \dots\right]
\eeq
where the sign ambiguity is irrelevant and can therefore be removed.

We need to enforce the boundary condition at $x=L$, thus obtaining the "quantization" condition
\beq
\sqrt{E_n} \int_{-L}^L \sqrt{\rho(y)}  dy - \frac{1}{32 \sqrt{E_n}} \int_{-L}^L \frac{ 5 \rho'(y)^2-4 \rho(y) \rho''(y)}{\rho(y)^{5/2}} 
dy = n\pi \ .
\eeq

Although this equation may be solved exactly it is convenient to assume
\beq
\sqrt{E_n} =\sum_{n=0}^\infty \alpha_n (n\pi)^{1-\alpha} \ ,
\eeq
and find the constant coefficients $\alpha_n$:
\beq
\alpha_0 &=& \frac{1}{\int_{-L}^L \sqrt{\rho(y)}  dy} \equiv \frac{1}{\sigma(L)}  \ , \\
\alpha_1 &=& 0 \\
\alpha_2 &=&  \frac{1}{32} \int_{-L}^L \frac{- 5 \rho'(y)^2+4 \rho(y) \rho''(y)}{\rho(y)^{5/2}} dy \\
\dots &=& \dots \nonumber 
\eeq

Therefore:
\beq
E_n \approx \frac{n^2\pi^2}{\sigma(L)^2} + \frac{1}{\sigma(L)} \int_{-L}^L \frac{- 5 \rho'(y)^2+4 \rho(y) \rho''(y)}{16 \rho(y)^{5/2}} dy 
+\dots \eeq

This expression agrees with the one found with WKBPT.

\section*{Acknowledgements}
We would like to acknowledge useful conversations with G. Morello and S.Rubatto.
The author acknowledges support of Conacyt through Sistema Nacional de Investigadores (SNI).

\end{document}